\documentstyle[12pt,epsfig]{article}
\newcommand{\be}{\begin{equation}}
\newcommand{\ee}{\end{equation}}
\newcommand{\bea}{\begin{eqnarray}}
\newcommand{\eea}{\end{eqnarray}}

\newcommand{\beq}{\begin{equation}}
\newcommand{\eeq}{\end{equation}}

\newcommand{\nn}{\nonumber}

\def\la{\mathrel{\mathpalette\fun <}}

\def\fun#1#2{\lower3.6pt\vbox{\baselineskip0pt\lineskip.9pt
\ialign{$\mathsurround=0pt#1\hfil##\hfil$\crcr#2\crcr\sim\crcr}}}

\begin{document}

\title{Quark--antiquark
states and their radiative transitions in terms of the spectral
integral  equation.\\ {\Huge III.} Light mesons.}
\author{V.V. Anisovich, L.G. Dakhno, M.A. Matveev,\\ V.A. Nikonov,
and A.V. Sarantsev}

\date{\today}
\maketitle

\begin{abstract}
We continue the investigation of mesons in terms
of the  spectral integral  equation initiated
before \cite{bb,cc} for the $b\bar b$ and $c\bar c$ systems: in this
paper we consider the light-quark ($u, d,s$) mesons with masses
$M\le 3$ GeV. The calculations have been performed for the mesons lying
on linear trajectories in the $(n,M^2)$-planes, where $n$  is the
radial quantum number. Our consideration relates to the $q\bar q$
states with one component in the flavor space, with the quark and
antiquark
 masses equal to each other, such as
 $\pi(0^{-+})$, $\rho(1^{--})$, $\omega(1^{--})$, $\phi(1^{--})$,
$a_0(0^{++})$, $a_1(1^{++})$, $a_2(2^{++})$,
$b_1(1^{+-})$, $f_2(2^{++})$, $\pi_2(2^{-+})$, $\rho_3(3^{--})$,
$\omega_3(3^{--})$, $\phi_3(3^{--})$, $\pi_4(4^{-+})$ at  $n\le
6$. We obtained the wave functions and mass values of mesons lying on
these trajectories. The corresponding trajectories are linear, in
agreement with data. We have calculated the two-photon decays $\pi\to
\gamma\gamma$, $a_0(980)\to \gamma\gamma$, $a_2(1320)\to \gamma\gamma$,
$f_2(1285)\to \gamma\gamma$, $f_2(1525)\to \gamma\gamma$
and radiative transitions $\rho\to\gamma\pi$, $\omega\to\gamma\pi$,
that agree qualitatively
with the experiment. On this basis, we extract the singular part of the
interaction amplitude, which corresponds to the so-called "confinement
interaction". The description of the data requires the presence of the
strong $t$-channel singularities for both scalar and vector exchanges.
\end{abstract}

\section{Introduction}

The present paper continues the investigation of the quark--antiquark
states within the  spectral integral equation:
 before, we investigated the  $b\bar  b$ \cite{bb} and $c\bar c$
 \cite{cc} systems, with the use of the information on the radiative
 decays \cite{raddecay}. In this line, the investigation of light
 quarks
$q\bar q$ is of a particular interest, for the highly excited
systems of light quarks are formed at large distances, where the
long-distance color forces, or "confinement forces", work. Just the
definition of the strong singularities of the amplitude, which provide
the quark confinement, is the main purpose of our
investigation.

Still, working with the quark--antiquark systems, we started with the
study of heavy quarks $(Q\bar Q=b\bar b,c\bar c)$, for such systems are
more simple for the investigation and it is more convenient to
develop the method using this pattern. Besides, the lowest
$Q\bar Q$ states may be considered, with rather good accuracy, in the
nonrelativistic approximation, and a number of papers, e.g., see
\cite{Hulth,Godfrey,Gupta,Lucha,NR} and references therein, is devoted
to  the study of
$Q\bar Q$ systems in the nonrelativistic approach.

The spectral integral method used in the analysis of the
quark--antiquark systems is a direct generalization of the dispersion
$N/D$ method \cite{chew} for the case of separable vertices. In the
framework of this method, we have analysed the two-nucleon systems and
their interaction with  electromagnetic field
\cite{deut}, in particular, the electric form factors of the deuteron
and deuteron photodisintegration amplitude  \cite{deut-AS}. In this
method, there was no problem with the description of the high-spin
particles. The method has been
generalized \cite{BS-YF} aiming to describe the quark--antiquark
systems. As a result, the equation was written for the
quark wave function, its form being similar to the Bethe--Salpeter
equation --- this was the reason to call it conventionally in
\cite{BS-YF} as the spectral integral Bethe - Salpeter equation.

The analyses of the light $q\bar q$ systems and heavy
$Q\bar Q$ quarkonia differ from one another
in certain respect, because the  corresponding
experimental data are different: in the $Q\bar Q$ systems the masses of
lowest states are known only, with an exception for the $1^{--}$
quarkonia ($\Upsilon$ and $\psi$)
where a long series of states was discovered in the $e^+e^-$
annihilation. At the same time, for the lowest
states there exists a rich set of data on radiative decays: $(Q\bar
Q)_{in}\to\gamma (Q\bar Q)_{out}$ and $(Q\bar Q)\to\gamma\gamma$. For
the light quark sector, there exists an abundant information on the
highly excited states with different $J^{PC}$, but rather poor one for
the radiative decays.

Despite the scarcity of data on radiative decays, we try to
apply the method to the study of light quarkonia, relying on our
knowledge of linear trajectories in the  $(n,M^2)$-plane ($n$ is the
radial quantum number of the $q\bar q$-meson with
mass $M$), that may somehow
compensate the lack of information on the wave functions. At the same
time, we present the predicted values for the two-photon decays,
that could stimulate their measurements.

The next Section is devoted to the presentation of the results. We do
not provide technical details of the calculations; all the calculation
machinery was explained in \cite{bb,raddecay}. In Conclusion,
we summarize our results.

\section{The results of calculations}

In this paper, we  study the systems with a single flavor
component. First, they are the systems with the unity isospin,
$(1,J^{PC})$. Second, among the systems with zero isospin,
$(0,J^{PC})$,  there
are also one-component states, $s\bar s$ or $n\bar n= (u\bar u+d\bar
d)/\sqrt{2}$, which are considered in this article as well: they are
$\phi$ and $\omega$ mesons, $\phi(1^{--}),\phi_3(3^{--})$ and
$\omega(1^{--}),\omega_3(3^{--})$. Besides, in the $f_2(2^{++})$-mesons
at $M\la 2400$ MeV the components $n\bar n$ and $s\bar s$ are separated
with a good accuracy \cite{f2}; below, all the $f_2$-mesons are assumed
to be pure flavor states.

Considering the $\pi$ trajectory [$\pi(140),\pi(1300),\pi(1800),
\pi(2070),\pi(2360$], we fix our attention on the excited states
$\pi(1300),\pi(1800),\pi(2070),\pi(2360)$. As concerns the lightest pion
  $\pi(140)$,
this particle  is a singular state in many respects, and we intend to
get qualitative agreement with data only (a good quantitative
description is beyond the scope of the present approach).

Generally speaking, the $1^{--},2^{++},3^{--}$ states are the mixtures
of waves with different angular momenta. However, the investigation of
the bottomonium and charmonium states \cite{bb,cc} shows us that the
angular momentum is a good quantum  number, so these states can be
described by one-component wave functions. Here, in the description of
the light quarks, we use as an ansatz the one-component wave functions
for $q\bar q$ systems.

We investigate the mesons with the masses $\la 3000$ MeV, they are
characterized by the following wave functions (see \cite{bb},
Section 2.1, for more detail):
\bea  \label{1}
\begin{tabular}{l|l|l}
              $L=0$ & $0^{-+}$ & $i\gamma_5\psi^{(0,0,0)}_n (k^2)$ \\
(dominant $S$-wave) & $1^{--}$ & $\gamma_\mu ^\perp\psi^{(1,0,1)}_n (k^2)$ \\ \hline
                    & $0^{++}$ & $ m\,\psi^{(1,1,0)}_n (k^2)$ \\
              $L=1$ & $1^{++}$ & $\sqrt{3/2s}\cdot i\,\varepsilon_{\gamma P k\mu}\psi^{(1,1,1)}_n (k^2)$ \\
(dominant $P$-wave) & $2^{++}$ & $\sqrt{3/4}\cdot \left [k_{\mu_1}\gamma^\perp_{\mu_2}+k_{\mu_2}\gamma^\perp_{\mu_1}-\frac{ 2}{3}\, \hat k g^\perp_{\mu_1 \mu_2}\right ]\psi^{(1,1,2)}_n (k^2)$ \\
                    & $1^{+-}$ & $\sqrt 3\, i\gamma_5 k_{\mu}\psi^{(0,1,1)}_n (k^2)$ \\  \hline
                    & $1^{--}$ & $3/\sqrt{ 2}\cdot \left [k_\mu\hat k-\frac13 k^2\gamma^{\perp}_\mu\right ]\psi^{(1,2,1)}_n (k^2)$ \\
              $L=2$ & $2^{--}$ & $\sqrt{20/9s}\cdot\,i\,\varepsilon_{\gamma P k\alpha}Z^{(2)}_{\mu_1\mu_2,\alpha}(k_\perp)\psi^{(1,2,2)}_n (k^2)$ \\
(dominant $D$-wave) & $3^{--}$ & $\sqrt{6/5}\cdot\gamma_\alpha Z^{(2)}_{\mu_1\mu_2\mu_3,\alpha}(k_\perp)\psi^{(1,2,3)}_n (k^2)$ \\
                    & $2^{-+}$ & $\sqrt{10/3}\cdot i\gamma_5 X^{(2)}_{\mu_1\mu_2}(k_\perp) \psi^{(0,2,2)}_n (k^2)$ \\ \hline
                    & $2^{++}$ & ${\sqrt 2}\cdot \gamma_\alpha X^{(3)}_{\mu_1\mu_2\alpha}(k_\perp)\psi^{(1,3,2)}_n (k^2)$ \\
              $L=3$ & $3^{++}$ & $\sqrt{21/10s}\cdot i\,\varepsilon_{\gamma P k\alpha}Z^{(2)}_{\mu_1\mu_2\mu_3,\alpha}(k_\perp)\psi^{(1,3,3)}_n (k^2)$ \\
(dominant $F$-wave) & $4^{++}$ & $\sqrt{36/35}\cdot\gamma_\alpha Z^{(3)}_{\mu_1\mu_2\mu_3\mu_4,\alpha}(k_\perp)\psi^{(1,3,4)}_n (k^2)$ \\
                    & $3^{+-}$ & $\sqrt{14/5}\cdot i\gamma_5 X^{(3)}_{\mu_1\mu_2\mu_3}(k_\perp) \psi^{(0,3,3)}_n (k^2)$ \\  \hline
                    & $3^{--}$ & $\sqrt{8/7}\cdot \gamma_\alpha X^{(4)}_{\mu_1\mu_2\mu_3\alpha}(k_\perp)\psi^{(1,4,3)}_n (k^2)$ \\
              $L=4$ & $4^{--}$ & $\sqrt{288/175s}\cdot\,i\,\varepsilon_{\gamma P k\alpha}Z^{(4)}_{\mu_1\mu_2\mu_3\mu_4,\alpha}(k_\perp)\psi^{(1,4,4)}_n (k^2)$ \\
(dominant $G$-wave) & $5^{--}$ & $\sqrt{16/35}\cdot\gamma_\alpha Z^{(2)}_{\mu_1\mu_2\mu_3\mu_4\mu_5,\alpha}(k_\perp)\psi^{(1,4,5)}_n (k^2)$ \\
                    & $4^{-+}$ & $\sqrt{81/35}\cdot i\gamma_5 X^{(4)}_{\mu_1\cdot\mu_4}(k_\perp)\psi^{(0,4,4)}_n (k^2)$ .
\end{tabular}
\eea
We characterize the group of mesons by the index $L$.
Recall that the index $L$ does not select a pure angular momentum state,
for example, the wave function $\gamma_\mu ^\perp\psi^{(1,0,1)}_n (k^2)$
given in (\ref{1}) for $(1^{--},L=0)$-system, being dominantly the
$S$-wave state, contains an admixture of the $D$-wave.

As our calculations show, the ansatz (\ref{1}) works well for the
considered mesons.

\subsection{Short-range interactions and confinement}

As compared to the approach applied in
\cite{bb,cc}, now there is one more simplification, which  relates
to the choice of the interaction: here we consider the instantaneous
interactions only. Such interaction worked well for the $b\bar b$ and
$c\bar c $ systems \cite{bb,cc}, hence it would be resonable to start
the investigation with similar hypothesis for the light quarks.

Recall that for the loosely bound states the equations to be solved are
written in the momentum representation. Below, however, to be more
illustative we present interaction amplitudes in the coordinate space.

As before \cite{bb,cc}, we use two types of the
$t$-channel interactions: scalar,
$(I\otimes I)$, and vector, $(\gamma_\nu\otimes\gamma_\mu)$, exchanges.
We classify the interactions as being short-range,
\be
\label{pot_2}
V_{s-r}(r)\ =\ a+c\,e^{-\mu_c r}+d\,\frac{e^{-\mu_d r}}{r}\ ,
\ee
and long-range (or confinement) ones:
\be
\label{pot_3}
V_{conf}(r)\ =\ b\,r\ .
\ee
In the momentum representation, the interactions
(\ref{pot_2}), (\ref{pot_3}) are written as the sets of the
$t$-channel poles:
\be
\label{pot_4}
I_N(t_\perp ,\mu)\ =\ \frac{4\pi(N+1)!}{(\mu^2-t_\perp)^{N+2}}
\sum^{N+1}_{j=0}(\mu+\sqrt{t_\perp})^{N+1-j}(\mu-\sqrt{t_\perp})^j\ .
\ee
We determine $t_\perp=(k^\perp_1-k'^\perp_1)_\mu
(-k^\perp_2+k'^\perp_2)_\mu$, where momenta $k_i$ and $k'_i$ refer to
quarks in the initial and intermediate states, here
 $(k^\perp_i\ ,k_1+k_2)=0$ and $(k'^\perp_i\ ,k'_1+k'_2)=0$ (for more
 detail see \cite{bb}). As is easy to see, the point $t_\perp=0$ is not
 singular: the right-hand side of (\ref{pot_4}) is the sum of pole
 terms $1/(\mu^2-t_\perp)^i$.

In the center-of-mass system, the pole functions $I_N(t_\perp)$
correspond to the right-hand side terms in (\ref{pot_2}),
(\ref{pot_3}) as follows:
\be
\label{pot_5}
I_N(t_\perp)\to r^N\,e^{-\mu r}\ ,
\ee
where the constant term in
(\ref{pot_2}) is written as a limit
$a\,I_0(t_\perp\ ,\mu_{const}\to 0)$ and the confinement term
(\ref{pot_3}) as $ b\, I_1(t_\perp\,\mu_{linear}\to 0)$.
Actually, we choose for the calculation
$\mu_{constant}$ and $\mu_{linear}$ of the order of $1-10$ MeV (it was
checked that the solutions for  $n\le 6$ are stable, when
 $\mu_{constant}$ and  $\mu_{linear}$ vary in such interval).

We fit to the $(n,M^2)$-tralectories separately for the states with
different $L$'s; but still we assume that the leading
(confinement) singularity is common for all  states (i.e., $b$ in
(\ref{pot_3}) is universal for all the $L$'s) while the short-range
interactions may be different for different $L$. Here, we take leave
of the pure potential approach used for the $b\bar b$ and $c\bar c$
systems in solutions $I$, $II$.
 Here, in fact, we adopt for the short-range interactions the
ideology of the dispersion relation $N/D$-method where the $N$-functions
for each wave may be considered independently.

So, we project $V_{s-r}(r)$ on the states with fixed $L$,
\be
\label{pot_7}
\langle L|V_{s-r}(r)|L\rangle\ ,
 \ee
and fit to each group of mesons with this short-range amplitude,
which is different for different $L$.

The fit gave us the following parameters for
$L=1,2,3,4$ (all values in GeV):
\be
\begin{tabular}{ccccccccccc}
Interaction & Wave & $a$  &  $b$ & $c$ & $\mu_c$ & $d$ & $\mu_d$ \\
\hline
$({\rm I} \otimes {\rm I})$ &
\begin{tabular}{c}
$L=0$ \\  $L=1$ \\ $L=2$  \\  $L=3$ \\ $L=4$ \\
\end{tabular}
&
\begin{tabular}{c}
                   -2.860 \\ -0.398 \\  8.407 \\ -0.281 \\ -1.912 \\
\end{tabular}
&
\begin{tabular}{c}
                    0.150 \\  0.150 \\  0.150 \\  0.150 \\  0.150 \\
\end{tabular}
&
\begin{tabular}{c}
                    5.057 \\  5.362 \\  6.866 \\  5.243 \\  3.8574 \\
\end{tabular}
&
\begin{tabular}{c}
                    0.410 \\  0.410 \\  0.110 \\  0.110 \\  0.010 \\
\end{tabular}
&
\begin{tabular}{c}
                    0.221 \\ -2.270 \\ -1.250 \\ -32.507 \\-3.3175\\
\end{tabular}
&
\begin{tabular}{c}
                    0.410 \\  0.210 \\  0.210 \\  0.410 \\  0.110 \\
\end{tabular}
\\ \hline
$(\gamma_{\mu} \otimes \gamma_{\mu})$&
\begin{tabular}{c}
$L=0$ \\  $L=1$ \\  $L=2$  \\  $L=3$ \\ $L=4$ \\
\end{tabular}
&
\begin{tabular}{c}
                     0.180 \\  0.971 \\  1.804 \\  1.239 \\  1.548 \\
\end{tabular}
&
\begin{tabular}{c}
                    -0.150 \\ -0.150 \\ -0.150 \\ -0.150 \\ -0.150 \\
\end{tabular}
&
\begin{tabular}{c}
                     0.060 \\ -1.888 \\ -2.135 \\ -12.823 \\-2.5458 \\
\end{tabular}
&
\begin{tabular}{c}
                     0.610 \\  0.610 \\  0.610 \\  0.710 \\  0.210 \\
\end{tabular}
&
\begin{tabular}{c}
                     0.656 \\  0.664 \\  0.405 \\  0.588 \\  0.536 \\
\end{tabular}
&
\begin{tabular}{c}
                     0.010 \\  0.010 \\  0.010 \\  0.010 \\  0.010 \\
\end{tabular}
\\ \hline
\end{tabular}
\ee
The fit requires the confinement
terms $V_{conf}(r)$ of Eq. (\ref{pot_3})   for both  scalar,
 $(I\otimes I)$,  and vector,
$(\gamma_\mu\otimes\gamma_\mu)$, exchanges, and the slopes of the
potentials  turn out to be approximately the same: $b_S\simeq -b_V$. In
the final fit, we fix the slopes to be equal thus obtaining
$b_S=|b_V|=0.15$ GeV$^2$.

The one-gluon exchange coupling $\alpha_s=\frac34 d_V$ is of the same
order for all $L$'s, $\alpha_s\sim 0.4$ --- this value looks quite
reasonable and agrees with other estimates, see, for example,
\cite{ryskin}.

We put for the masses of the constituent quarks $m_u=m_d=400$ MeV and
$m_s=500$ MeV.

\subsection{Masses and mean square radii of mesons with
$L\leq 4$}

Here, we present the resuls of calculations for the masses and mean
square radii of the mesons with $L=1,2,3,4$. The mean square radius of
quark--antiquark system is an important characreristics, which for
highly excited states is related to the confinement forces;
by considering them, it would be to useful keep in mind that for
the pion $R^2\simeq 10$ GeV$^{-2}$.

By listing the experimentally observed states, we follow
\cite{PDG,ufn04,f2,longacre}.

\subsubsection{Mesons of the $(L=0)$-group}

The calculation of the $(L=0)$-states leads to the following masses
(in MeV)
and mean squared radii ($R^2$ in GeV$^{-2}$) for the $(10^{-+},L=0)$ and
$(11^{--},L=0)$ mesons:
 \begin{equation}
\begin{tabular}{l|lll|lll}
n & Meson       & Mass & $R^2$ & Meson        & Mass & $R^2$ \\ \hline
1 & $\pi(140)$  &  546 & 12.91 & $\rho(770)$  &  778 & 12.77 \\
2 & $\pi(1300)$ & 1309 & 33.94 & $\rho(1450)$ & 1473 & 12.34 \\
3 & $\pi(1800)$ & 1771 & 62.26 & $\rho(1830)$ & 1763 & 18.08 \\
4 & $\pi(2070)$ & 2009 & 29.02 & $\rho(2110)$ & 2158 & 45.71 \\
5 & $\pi(2360)$ & 2429 & 29.86 &  ---         & 2363 & 60.30 \\
6 &   ---       & 3075 & 25.07 &  ---         & 2675 & 26.17\ .
\end{tabular}
\label{m_S_1}
\end{equation}
For isoscalar $(01^{--},L=0)$ mesons, we assume $\omega$ and $\phi$ to
be pure flavor states
($\omega=(u\bar u +d\bar d)/\sqrt 2 $ and $\phi
=s\bar s$) and obtain:
\begin{equation}
\begin{tabular}{l|lll|lll}
n & Meson          & Mass & $R^2$ & Meson        & Mass & $R^2$ \\ \hline
1 & $\omega(780)$  &  778 & 12.77 & $\phi(1020)$ &  938 & 16.20\\
2 & $\omega(1420)$ & 1473 & 12.34 & $\phi(1657)$ & 1541 & 21.29 \\
3 & $\omega(1800)$ & 1763 & 18.08 &  ---         & 1907 & 22.18 \\
4 & $\omega(2150)$ & 2158 & 45.71 &  ---         & 2327 & 31.88 \\
5 & ---            & 2363 & 60.30 &  ---         & 2601 & 78.03 \\
6 &   ---          & 2675 & 26.17 &  ---         & 2757 & 23.84 \ .
\end{tabular}
\label{m_S_2}
\end{equation}
The latest measurement of the mass $\phi(1680)$ provided the value
$1623\pm20$ MeV \cite{akhmetshin}, in the close agreement with the
calculations.

\subsubsection{Mesons of the $(L=1)$-group}

In the isovector sector, we obtain
 for $(10^{++},L=1)$ and $(11^{++},L=1)$ mesons:
\bea
\begin{tabular}{l|lll|lll}
n & Meson       & Mass  & $R^2$ & Meson       & Mass  & $R^2$ \\ \hline
1 & $a_0( 980)$ & 1035 &  7.19 & $a_1(1230)$ & 1151 &  6.88 \\
2 & $a_0(1515)$ & 1496 & 13.57 & $a_1(1640)$ & 1562 & 13.67 \\
3 & $a_0(1830)$ & 1884 & 21.63 & $a_1(1970)$ & 1923 & 21.95 \\
4 & $a_0(2120)$ & 2208 & 30.72 & $a_1(2270)$ & 2231 & 42.81 \\
5 & ---         & 2488 & 42.78 & ---         & 2305 & 48.03 \\
6 & ---         & 2777 & 39.98 & ---         & 2682 & 28.66\ ,
\end{tabular}
\label{massesP1}
\eea
and  for $(12^{++},L=1)$ and $(11^{+-},L=1)$ ones:
\bea
\begin{tabular}{l|lll|lll}
n & Meson       & Mass & $R^2$ & Meson       & Mass & $R^2$ \\ \hline
1 & $a_2(1320)$ & 1356 &  7.08 & $b_1(1235)$ & 1168 &  7.01 \\
2 & $a_2(1660)$ & 1641 & 13.89 & $b_1(1640)$ & 1567 & 13.67 \\
3 & $a_2(1950)$ & 1963 & 22.05 & $b_1(1970)$ & 1928 & 21.73 \\
4 & $a_2(2255)$ & 2260 & 31.76 & $b_1(2210)$ & 2240 & 31.03 \\
5 & ---         & 2517 & 43.70 & ---         & 2548 & 34.70 \\
6 & ---         & 2810 & 37.82 & ---         & 2927 & 32.32\ .
\end{tabular}
\label{massesP1}
\eea
The fitting to the ($02^{++},L=1$) mesons is performed separarely for
$n\bar n$ and $s\bar s$ systems: the analysis of the decay couplings
$f_2 \to K\bar K, \pi\pi,\eta\eta,\eta\eta'$ \cite{f2}
argues that $f_2$-mesons at $M\leq 2400$ MeV
are nealy pure $n\bar n$ or $s\bar s$ states.
Analogous arguments follow from the data on $\pi^-p\to \phi\phi p$
\cite{longacre}. We have:
\bea
\begin{tabular}{l|lll|lll}
n & Meson $(n\bar n)$       &Mass   & $R^2$ & Meson $(s\bar s)$
 & Mass   & $R^2$ \\ \hline
1 & $f_2(1285)$ & 1356 &  7.08 & $f_2(1525)$ & 1608 &  6.52 \\
2 & $f_2(1580)$ & 1641 & 13.89 & $f_2(1755)$ & 1855 & 11.88 \\
3 & $f_2(1920)$ & 1963 & 22.05 & $f_2(2120)$ & 2162 & 18.76 \\
4 & $f_2(2240)$ & 2260 & 31.76 & $f_2(2410)$ & 2454 & 26.27 \\
5 & ---         & 2516 & 43.70 & ---         & 2731 & 36.42 \\
6 & ---         & 2809 & 37.82 & ---         & 2990 & 40.62 \, .
\end{tabular}
\label{massesP2}
\eea
Let us stress that the fit gives us comparatively small values for
$R^2[f_2(1285)]$ and $R^2[f_2(1525)]$,
($\simeq 7$ GeV$^{-2}$): just such small values are
requiered by the $\gamma\gamma$ decays of these tensor mesons,
see \cite{f2-gg}.

\subsubsection{Mesons of the $(L=2)$-group}

We have the
masses and mean square radii
(in Mev)  for  the
dominant ($^1D_2,I_{q\bar q}=1$) and ($^3D_1,I_{q\bar q}=1$) systems:
\bea
&&\begin{tabular}{l|lll|lll}
n & Meson         & Mass & $R^2$ & Meson        & Mass & $R^2$ \\ \hline
1 & $\pi_2(1670)$ & 1700 &  5.81 & $\rho(1690)$ & 1701 &  8.14 \\
2 & $\pi_2(2005)$ & 1937 & 11.53 & $\rho(2010)$ & 1992 & 15.26 \\
3 & $\pi_2(2245)$ & 2348 & 16.44 & $\rho(2260)$ & 2212 & 31.44 \\
4 & $\pi_2(2510)$ & 2637 & 22.98 & ---          & 2515 & 30.33 \\
5 & ---           & 2914 & 28.70 & ---          & 2743 & 11.24 \, ,
\end{tabular}
\eea
for               the
dominant ($^3D_3,I_{q\bar q}=1$) and ($^3D_1,I_{q\bar q}=0$) systems:
\bea
&&\begin{tabular}{l|ll|ll}
n & Meson          & Mass   & Meson          & Mass \\ \hline
1 & $\rho_3(1688)$ & 1671 & $\omega_1(1640)$ & 1701 \\
2 & $\rho_3(1982)$ & 1987 & $\omega_1(1920)$ & 1992 \\
3 & $\rho_3(2250)$ & 2376 & $\omega_1(2295)$ & 2212 \\
4 & ---            & 2705 & ---              & 2515 \\
5 & ---            & 2991 & ---              & 2743 \, ,
\end{tabular}
\eea
and for              the
dominant ($^3D_3,I_{q\bar q}=0$) and ($^3D_1s\bar s$) systems:
\bea
&&\begin{tabular}{l|ll|ll}
n & Meson            & Mass & Meson          & Mass \\ \hline
1 & $\omega_3(1670)$ & 1671 & $\phi_3(1850)$ & 1850 \\
2 & $\omega_3(1960)$ & 1987 & $\phi_3(2150)$ & 2150 \\
3 & $\omega_3(2295)$ & 2376 & $\phi_3(2450)$ & 2450 \\
4 & ---              & 2705 & $\phi_3(2640)$ & 2514 \\
5 & ---              & 2991 & ---            & 2617 \  .
\end{tabular}
\label{massesD}
\eea

\subsubsection{Mesons of the $(L=3)$-group}

We calculate the following $q\bar q$-mesons with dominant $F$-wave:
in the $(I=1)$ sector,
\bea
&&\begin{tabular}{l|lll|ll}
n & Meson       & Mass & $R^2$ & Meson       & Mass \\ \hline
1 & $a_2(2030)$ & 2019 & 10.88 & $a_3(2030)$ & 2062 \\
2 & $a_2(2310)$ & 2263 & 22.42 & $a_3(2275)$ & 2314 \\
3 & ---         & 2460 & 29.63 & ---         & 2585 \\
4 & ---         & 2847 & 37.00 & ---         & 2938 \\
5 & ---         & 3360 & 38.11 & ---         & 3390 \, ,
\end{tabular}
\eea
and
\bea
&&\begin{tabular}{l|ll|ll}
n & Meson       & Mass & Meson       & Mass \\ \hline
1 & $b_3(2020)$ & 2013 & $a_4(2005)$ & 2018 \\
2 & $b_3(2245)$ & 2291 & $a_4(2255)$ & 2333 \\
3 & $b_3(2520)$ & 2538 & ---         & 2493 \\
4 & $b_3(2740)$ & 2706 & ---         & 2659 \\
5 & ---         & 3065 & ---         & 3059 \, ,
\end{tabular}
\eea
and in the $(I=0)$ sector,
\bea
&&\begin{tabular}{l|ll|ll}
n & Meson $(n\bar n)$ & Mass & Meson $(s\bar s)$& Mass  \\ \hline
1 & $f_2(2020)$ & 2018 & $f_2(2340)$ & 2315  \\
2 & $f_2(2300)$ & 2262 & ---         & 2498  \\
3 & ---         & 2460 & ---         & 2770  \\
4 & ---         & 2846 & ---         & 3136  \\
5 & ---         & 3360 & ---         & 3591  \ .
\end{tabular} \label{massesF} \eea

\subsubsection{Mesons of the $(L=4)$-group}

In the $(L=4)$-group, we calculate the lightest mesons:
\begin{equation}
\begin{tabular}{l|ll|ll}
n & Meson          & Mass & Meson         & Mass \\ \hline
1 & $\rho_3(2260)$ & 2252 & $\pi_4(2250)$ & 2257 \\
2 & ---            & 2482 & ---           & 2516 \\
3 & ---            & 2746 & ---           & 2842 \\
4 & ---            & 3131 & ---           & 3268 \\
5 & ---            & 3607 & ---           & 3760\ .
\end{tabular}
\label{massesG}
\end{equation}

\subsection{Trajectories in the $(n,M^2)$-planes}

\begin{figure}
\centerline{\epsfig{file=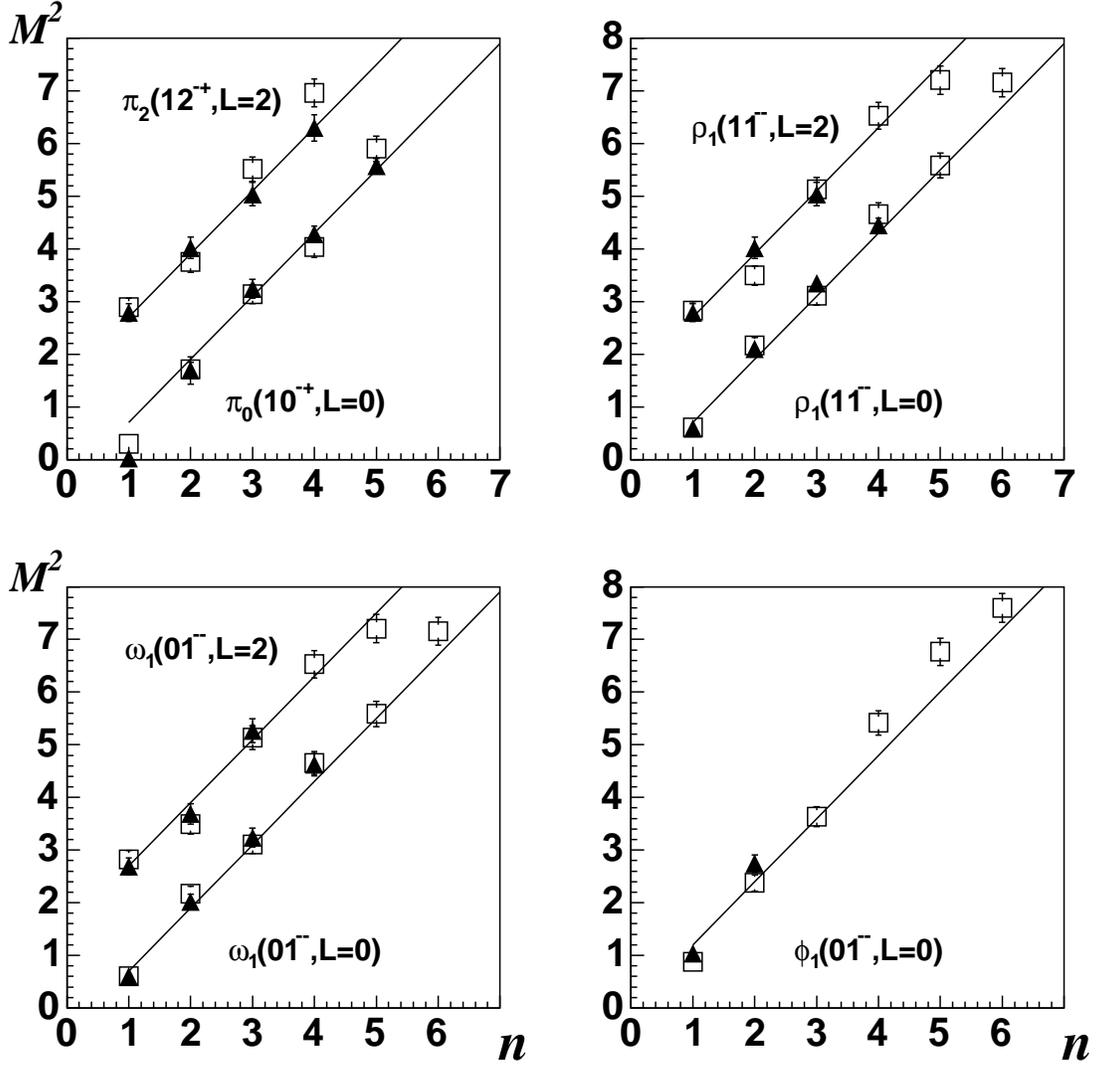,width=15cm}}
\caption{The $(L=0)$ and $(L=2)$ trajectories on the $(n,M^2)$-planes,
see Eqs. (20) and (21). Open squares stand for the calculated masses.
Thin lines represent linear trajectories, Eq. (26), with $\mu=1.2$
GeV$^2$.}
\end{figure}

\begin{figure}
\centerline{\epsfig{file=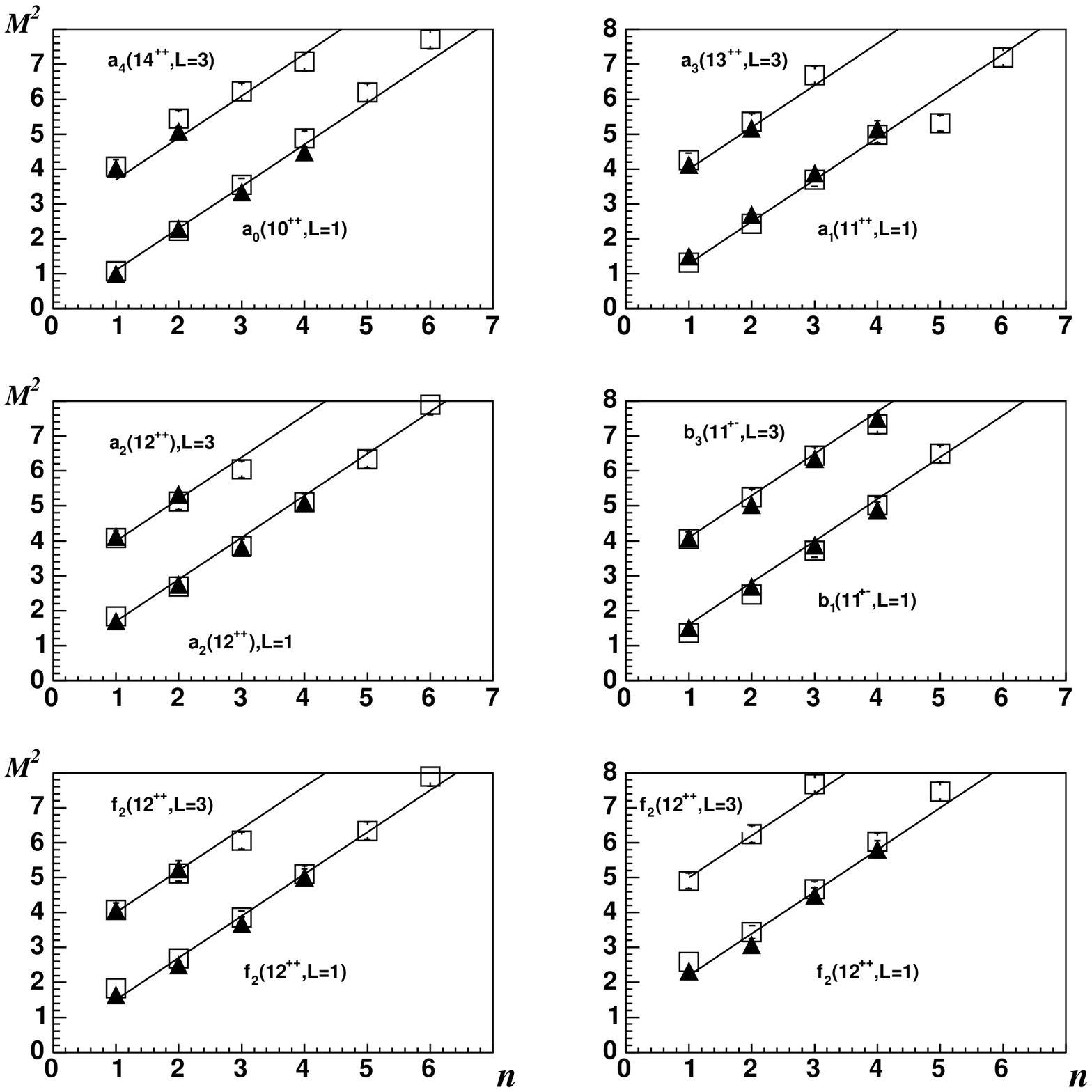,width=15cm}}
\caption{The $(L=1)$ and $(L=3)$ trajectories on the $(n,M^2)$-planes,
see Eqs. (22) and (23). The notations are as in Fig. 1.}
\end{figure}

\begin{figure}
\centerline{\epsfig{file=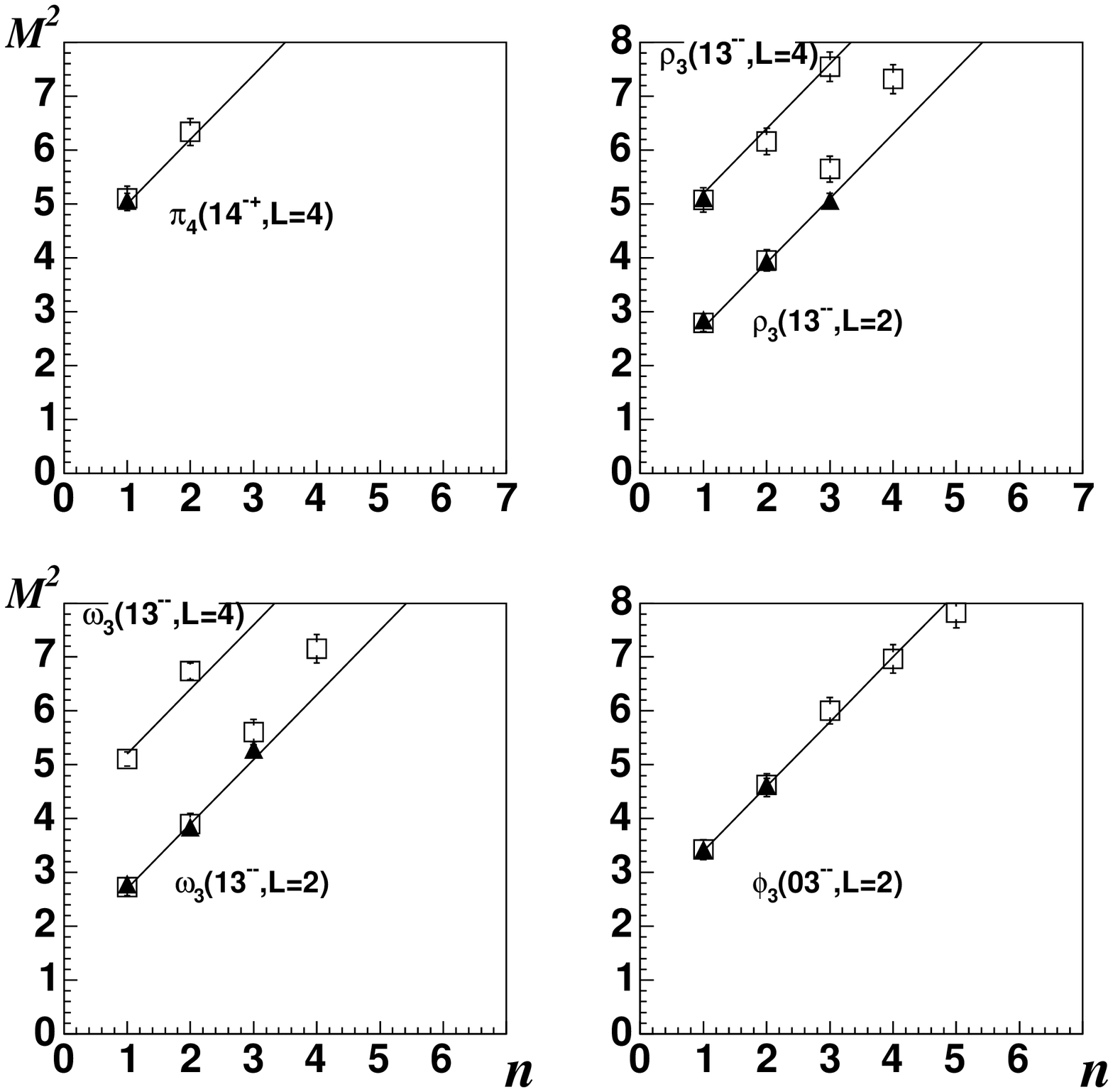,width=15cm}}
\caption{The $(L=2)$ and $(L=4)$ trajectories on the $(n,M^2)$-planes,
see Eqs. (24) and (25). The notations are as in Fig. 1.}
 \end{figure}

In Figs. 1,2,3, one may see the $q\bar q$-trajectories in the $(n,M^2)$
planes (bold numbers stand for the predicted states).

In Fig. 1, we show the trajectories for the $(L=0)$-group:
\bea
\label{fig1a}
&&[\pi(10^{-+},L=0)]= [\pi(140),  \pi(1300), \pi(1800), \pi(2070),      \pi(2360),  \pi({\bf 3075})], \\ \nn
&&[\rho(11^{--},L=0)]=[\rho(770),\rho(1450),\rho(1830),\rho(2110),\rho({\bf 2363}),\rho({\bf 2675}) ], \\ \nn
&&[\omega(01^{--},L=0)]=[\omega(780),\omega(1420),\omega(1800), \omega(2150),\omega({\bf 2363}),\omega({\bf 2675})], \\ \nn
&&[\phi(01^{--},L=0)]=[\phi(1020),\phi(1657),\phi({\bf 1907}),\phi({\bf 2327}),\phi({\bf 2601}),\phi({\bf 2757})]\ ,
\eea
and $(L=2)$-group:
\bea
\label{fig1b}
&&[\pi(12^{-+},L=2)]=[\pi(1670),\pi(2005),\pi(2245),\pi(2510)],\\ \nn
&&[\rho(11^{--},L=2)]=[\rho(1690),\rho(2010),\rho(2260),\rho({\bf 2515}), \rho({\bf 2743})], \\ \nn
&&[\omega(01^{--},L=2)]=[\omega(1640),\omega(1920),\omega(2295), \omega({\bf 2515}),\omega({\bf 2743})]\ .
\eea

In Fig. 2, we demonsrate the $L=1$ and $L=3$ trajectories. We have for
$L=1$:
\bea
\label{fig2a}
&&[a(10^{++},L=1)]=[a_0(1000),a_0(1515),a_0(1830),a_0(2120)], \\ \nn
&&[a(11^{++},L=1)]=[a_1(1230),a_1(1640),a_1(1970),a_1(2270)], \\ \nn
&&[a(12^{++},L=1)]=[a_2(1308),a_2(1660),a_2(1950),a_2(2255)], \\ \nn
&&[b(11^{+-},L=1)]=[b_1(1235),b_1(1640),b_1(1970),b_1(2210)], \\ \nn
&&[f(02^{++},n\bar n,L=1)]=[f_2(1285),f_2(1580),f_2(1920),f_2(2240)], \\ \nn
&&[f(02^{++},s\bar s,L=1)]=[f_2(1525),f_2(1755),f_2(2120),f_2(2410)]\ ,
\eea
and for $L=3$:
\bea
\label{fig2b}
&&[a(12^{++},L=3)]=[a_2(2030),a_2(2310),a_2({\bf 2460}),a_2({\bf 2847})], \\ \nn
&&[a(10^{++},L=3)]=[a_3(2030),a_3(2275),a_3({\bf 2585}),a_3({\bf 2938})], \\ \nn
&&[a(11^{++},L=3)]=[a_4(2005),a_4(2255),a_4({\bf 2493}),a_4({\bf 2659})], \\ \nn
&&[b(11^{+-},L=3)]=[b_3(2020),b_3(2245),b_3(2520),b_3(2740)], \\ \nn
&&[f(02^{++},n\bar n,L=3)]=[f_2(2020),f_2(2290),f_2({\bf 2460}),f_2({\bf 2846})], \\ \nn
&&[f(02^{++},s\bar s,L=3)]=[f_2(2340),f_2({\bf 2498}),f_2({\bf 2770}),
         f_2({\bf 3136})].
\eea
The trajectories for the $L=2$ and $L=4$ groups are shown in Fig. 3:
\bea
\label{fig3a}
&&[\rho(13^{--},L=2)]=[\rho_3(1688),\rho_3(1982),\rho_3(2250),
         \rho_3({\bf 2705}),\rho_3({\bf 2991})], \\ \nn
&&[\omega(03^{--},L=2)]=[\omega_3(1640),\omega_3(1920),
         \omega_3(2295),\omega_3({\bf 2705}),
         \omega_3({\bf 2991})], \\ \nn
&&[\phi(03^{--},L=2)]=[\phi_3(1850),\phi_3(2150),\phi_3(2450),
         \phi_3(2640),\phi_3({\bf 2617})]\ ,
\eea
and
\bea
&&[\pi(14^{-+},L=4)]=[\pi_4(2250),\pi_4({\bf 2516}),\pi_4({\bf 2842}),
         \pi_4({\bf 3268})],
  \\ \nn
&&[\rho(13^{--},L=4)]=[\rho_3(2260),\rho_3({\bf 2482}),
         \rho_3({\bf 2746}),\rho_3({\bf 3131}),
         \rho_3({\bf 3607})], \\ \nn
&&[\omega(03^{--},L=4)]=[\omega_3({\bf 2252}),\omega_3({\bf 2482}),
         \omega_3({\bf 2746}),\omega_3({\bf 3131}),\omega_3({\bf 3607})]\ .
\eea
In
agreement with the observation  \cite{syst}, all trajectories are
linear, with a good accuracy:
\be \label{traj1} M^2=M^2_0+\mu^2(n-1),
\ee
and have a universal slope:
\be
\label{traj2}
\mu^2\simeq 1.2\quad {\rm GeV}^2.
\ee

\section{Radiative decays}

The main source of the information on the wave functions of the $q\bar
q$ states is the two-photon meson decays. The matter is that the
two-photon decay amplitude is the convolution of meson wave function
and the quark component of the photon wave function \cite{PR-g}. The
light-quark component of the photon is rather reliably determined at
present time \cite{YF-g}. In this way, we calculate the widths for the
two-photon decays as follows:
\begin{equation}
\begin{tabular}{l|lll}
$q\bar q$-state & Process & Data (keV) & Fit (keV) \\ \hline
$1^1S_0$ & $\pi (140)  \to\gamma\gamma$ & 0.007          & 0.005  \\
$1^1S_0$ & $\pi (1300) \to\gamma\gamma$ & ---            & 3.742  \\
$1^1S_0$ & $\pi (1800) \to\gamma\gamma$ & ---            & 8.466  \\
$1^3P_0$ & $a_0 (980) \to\gamma\gamma$ & 0.300$\pm$0.100 & 0.340  \\
$1^3P_0$ & $a_0 (1515) \to\gamma\gamma$ & ---            & 0.224  \\
$1^3P_0$ & $a_0 (1830) \to\gamma\gamma$ & ---            & 0.186  \\
$1^3P_2$ & $a_2 (1320) \to\gamma\gamma$ & 1.00$\pm$0.06  & 1.045  \\
$1^3P_2$ & $a_2 (1660) \to\gamma\gamma$ & ---            & 0.821  \\
$1^3P_2$ & $a_2 (1950) \to\gamma\gamma$ & ---            & 0.699  \\
$1^3P_2\,n\bar n$ & $f_2 (1285) \to\gamma\gamma$ & 2.71$\pm$0.25 & 2.946  \\
$1^3P_2\,n\bar n$ & $f_2 (1580) \to\gamma\gamma$ & ---           & 2.396  \\
$1^3P_2\,n\bar n$ & $f_2 (1920) \to\gamma\gamma$ & ---           & 1.971  \\
$1^3P_2\,s\bar s$ & $f_2 (1525) \to\gamma\gamma$ & 0.10$\pm$0.01 & 0.135  \\
$1^3P_2\,s\bar s$ & $f_2 (1755) \to\gamma\gamma$ & ---           & 0.118  \\
$1^3P_2\,s\bar s$ & $f_2 (2120) \to\gamma\gamma$ & ---           & 0.097  \\
\end{tabular}
\label{qq-gg}
\end{equation}
Concerning the measured widths, we have a good agreement with the
calculated magnitudes. One should pay attention to the proximity of the
calculated width $\Gamma(\pi^0\to \gamma\gamma)\simeq 0.005$ keV and
experimental value (in the calculation, the real mass of pion was taken
for the phase space). This proximity tells us that the calculated wave
function is close to the real one, despite a large difference between
real and calculated pion masses. The information  on the radiative
decays of vector mesons tells us the same:
\begin{equation}
\begin{tabular}{lll}
Process                             & Data (keV) & Fit (keV)   \\
$\rho^+(770)\to \gamma\pi^+(140)$   &  68$\pm$7  &  36 \\
$\omega (780)\to \gamma\pi^0(135)$ & 758$\pm$25 & 323 \\
\label{qq-gqq}
\end{tabular}
\end{equation}
One may see the qualitative agreement with data (the difference of
amplitudes is of the order of $40\%$, being slightly larger than in
$\pi^0\to \gamma\gamma$); still,  let us underline that the decays
$V\to \gamma P$ are determined by the $M1$ transitions, which are
sensitive to the presence of the small contributions initiated by the
anomalous magnetic moment, e.g., see the discussion in \cite{Az,Ger}).
One may think that the corrections to  the $\pi(140)$ mass and its
wave function can be easily reached in the standard way, with
instanton-induced interaction (e.g., see \cite{instanton,gerasyuta} and
regerences therein). From this point of view, typical are the results
obtained in \cite{gerasyuta}, where the bootstrap quark model was
considered for the three lowest meson nonets $^1S_0,^3S_0,^3P_0$.
Without instanton-induced interactions, the pion mass was obtained to
equal $\sim 500$ MeV, while the input of these forces in the
calculation made the pion mass to be near 140 MeV. But we are
not interested in the precise description of the pion,  not
dealing with this problem right now.

We have a good description of the available experimental data for
$V\to e^+e^-$:
\begin{equation}
\begin{tabular}{lllll}
Process                  & Data          & Fit   \\
$\rho(770) \to e^+e^-$   & 7.02$\pm$0.11 & 7.260 \\
$\rho(1450)\to e^+e^-$   & ---           & 3.280 \\
$\rho(1830)\to e^+e^-$   & ---           & 2.790 \\
$\rho(2110)\to e^+e^-$   & ---           & 2.431 \\
$\omega(780) \to e^+e^-$ & 0.60$\pm$0.02 & 0.776 \\
$\omega(1420)\to e^+e^-$ & ---           & 0.388 \\
$\omega(1800)\to e^+e^-$ & ---           & 0.326 \\
$\omega(2150)\to e^+e^-$ & ---           & 0.255 \\
$\phi(1020)\to e^+e^-$   & 1.27$\pm$0.04 & 1.353 \\
$\phi(1657)\to e^+e^-$   & ---           & 0.985 \\
\end{tabular}
\end{equation}
These decays are determined by the convolution of the vector meson wave
functions and the $\gamma\to q \bar q$ vertex, whose parameters were
found in \cite{YF-g}.

\section{Conclusion}

We have a good description of mesons treated as the
bound states of constituent quarks: the masses of mesons with
one flavor component lay on the linear trajectories in the
$(n,M^2)$-plane. Also we have a not bad coincidence of the measured and
calculated widths of the radiative decays, though one should underline a
scarcity of the available data.

The main purpose of our investigation of the light quark sector is to
determine the characteristics of the leading $t$-channel singularities
(confinement singularities or, in the language of potentials, the
confinement potentials $V_{conf} (r)\sim br$). Our solution requires
the scalar and vector $t$-channel exchanges; in the color space this
is an exchange of the quantum numbers $c=1+8$ (pure octet exchange is
also possible).

We have described the data assuming for the scalar and vector exchanges
 the "confinement singularity
couplings" to be equal to each other (we mean the equality  of
slopes for the scalar and vector confinement potentials: $b_S=|b_V|$).
We do not know to what extent these magnitudes may differ; for such a
study, a more abundand data are needed, first of all, the data on the
radiative decays. But one statement may be pronounced, that is, if
$b_V=0$, the description of data is violated. Note that the variant
with $b_V\ne 0$ was frequently discussed, e.g., see \cite{Lucha} and
references therein.

We pay a considerable attention to the obtained wave
functions. The matter is that, to reconstruct the interaction, the only
knowledge on the mass levels is not enough --- one should know
the wave functions of mesons (see the discussion in \cite{BS-YF}).
Because of that, a simultaneous presentation of the calculated levels
and their wave functions is absulely necessary for both understanding
the results and the verification of  predictions.

\section*{Acknowledgments}

We thank A.V. Anisovich, Y.I. Azimov, G.S. Danilov, I.T. Dyatlov,
L.N. Lipatov, V.Y. Petrov, H.R. Petry and M.G. Ryskin
for useful discussions.

This work was supported by the Russian Foundation for Basic Research,
project no. 04-02-17091.

\section*{Appendix. Wave function of the quark-antiquark states}

Tables 1--8 give us the $c_i(S,L,J;n)$ coefficients,
which determine the wave functions $\psi^{(S,L,J)}$
according to the following formula:
\bea
\nonumber
\label{a-1}
\psi^{(S,L,J)}_{(n)}(k^2)=e^{-\beta k^2}\sum\limits_{i=1}^9 c_i(S,L,J;n)
k^{i-1}\, ,
\eea
where $ k^2\equiv {\bf k}^2 $
(recall that $s=4m^2+4{\bf k}^2$). As in the previous papers
\cite{bb,cc}, we fix
the
fitting parameter $\beta=1.2 $ GeV$^{-2}$.
Normalization condition for $\psi^{(S,L,J)}_{(n)}(k^2)$ is presented in
\cite{bb}.
 In Figs. 4--24, we
demonstrate these wave functions.
\begin{table}
\caption{Constants $c_i(S,L,J;n)$ (in GeV units) for
the wave functions of mesons with $L=0$ }
\begin{center} {\scriptsize
\begin{tabular}{|r|r|r|r|r|r|r|}
\hline
& $\pi(1S)$ & $\pi(2S)$ & $\pi(3S)$ & $\pi(4S)$ & $\pi(5S)$ & $\pi(6S)$ \\
\hline
$i$ & $\psi_{1}^{(0,0,0)}$ & $\psi_{2}^{(0,0,0)}$ & $\psi_{3}^{(0,0,0)}$ & $\psi_{4}^{(0,0,0)}$  & $\psi_{5}^{(0,0,0)}$ & $\psi_{6}^{(0,0,0)}$  \\
\hline
 1 &        51.6431 &       132.0357 &      -349.9543 &       110.9733 &       110.9373 &        -0.9078 \\
 2 &       -75.4242 &     -3416.0503 &      5923.8110 &     -1026.4241 &     -1406.4444 &       375.4569 \\
 3 &      -786.2797 &     26717.9980 &    -38671.3217 &      2223.1070 &      5487.4805 &     -5136.3924 \\
 4 &      3369.5822 &    -97897.7802 &    130528.4044 &      2962.3226 &     -6767.6992 &     26313.9583 \\
 5 &     -5983.5800 &    197748.6733 &   -253088.3937 &    -18810.8623 &     -6068.5199 &    -66856.4436 \\
 6 &      5700.2596 &   -232791.3934 &    291304.4722 &     31139.4532 &     24696.4108 &     92630.1398 \\
 7 &     -2952.2781 &    155832.6476 &   -192356.5330 &    -24525.2288 &    -25534.2375 &    -69743.1180 \\
 8 &       694.6688 &    -49062.0292 &     60017.9674 &      8448.0874 &     10222.6874 &     23922.9142 \\
 9 &        12.5202 &      -621.4046 &       758.7418 &       136.3225 &       162.3905 &       366.0005 \\
10 &       -48.0334 &      3035.1343 &     -3694.9898 &      -640.7132 &      -805.2658 &     -1749.6336 \\
11 &        21.9467 &       856.9842 &     -1008.6265 &      -102.4872 &      -272.4831 &      -403.0727 \\
12 &        -3.5217 &      -520.0370 &       621.5647 &        85.9917 &       156.4406 &       275.8358 \\
13 &        -3.5146 &       -13.0445 &        10.2613 &        -0.6161 &        17.9941 &        10.0999 \\
14 &         0.7801 &       -22.6780 &        28.7625 &         4.4634 &         2.7902 &        11.2089 \\
\hline
\hline
& $\rho_1(1S)$ & $\rho_1(2S)$ & $\rho_1(3S)$ & $\rho_1(4S)$ & $\rho_1(5S)$ & $\rho_1(6S)$ \\
\hline
$i$ & $\psi_{1}^{(1,0,1)}$ & $\psi_{2}^{(1,2,1)}$ & $\psi_{3}^{(1,0,1)}$ & $\psi_{4}^{(1,2,1)}$  & $\psi_{5}^{(1,0,1)}$ & $\psi_{6}^{(1,2,1)}$  \\
\hline
 1 &        44.2644 &       -47.0541 &        34.4676 &       256.1470 &      -324.1004 &        88.8577 \\
 2 &       147.9287 &        96.4391 &       367.3542 &     -3816.4658 &      6084.9026 &     -2239.2771 \\
 3 &     -2576.7481 &      1694.4435 &     -6627.1107 &     21285.8878 &    -41681.2603 &     19191.2661 \\
 4 &     10145.9141 &     -8835.1303 &     31300.6081 &    -61891.6195 &    143324.3794 &    -78486.1587 \\
 5 &    -20331.5277 &     18954.3680 &    -72495.7682 &    106967.9533 &   -280751.9128 &    176402.4007 \\
 6 &     23805.7270 &    -21715.0204 &     95497.7091 &   -115547.6449 &    329759.5447 &   -232126.0612 \\
 7 &    -16569.8166 &     13585.9932 &    -73882.6198 &     77608.2218 &   -231221.1727 &    180168.7079 \\
 8 &      6338.4464 &     -3952.2967 &     31633.5665 &    -29980.2983 &     88892.1822 &    -76972.7416 \\
 9 &      -941.1781 &       119.3938 &     -5588.5913 &      4927.5676 &    -13196.2908 &     13344.7807 \\
10 &       -59.0451 &        26.4018 &      -333.1001 &       258.1260 &      -807.8113 &       812.7029 \\
11 &       -16.0783 &        88.7553 &        43.2902 &       -25.9820 &      -250.3708 &       -36.0072 \\
12 &         5.1228 &        -4.9922 &        23.4994 &       -20.3851 &        71.5941 &       -61.2100 \\
13 &         6.8825 &        -1.8124 &        41.0982 &       -31.6383 &        94.3662 &       -99.2026 \\
14 &         1.6305 &       -10.0019 &        -4.8100 &         4.6716 &        27.0834 &         2.6292 \\
\hline
\hline
& $\phi_1(1S)$ & $\phi_1(2S)$ & $\phi_1(3S)$ & $\phi_1(4S)$ & $\phi_1(5S)$ & $\phi_1(6S)$ \\
\hline
$i$ & $\psi_{1}^{(1,0,1)}$ & $\psi_{2}^{(1,0,1)}$ & $\psi_{3}^{(1,0,1)}$ &
$\psi_{4}^{(1,0,1)}$  & $\psi_{5}^{(1,0,1)}$ & $\psi_{6}^{(1,0,1)}$  \\ \hline
 1 &        38.1719 &        22.5404 &        18.7861 &      -155.5852 &       352.9271 &        -4.8085 \\
 2 &       295.3927 &       455.1981 &     -1230.3142 &      2157.2573 &     -6466.3709 &       465.4045 \\
 3 &     -3779.0011 &     -6708.1413 &     12765.6782 &    -10539.3998 &     44222.2695 &     -6120.4969 \\
 4 &     14301.6996 &     29467.5970 &    -55150.4290 &     24553.7589 &   -153801.5356 &     31536.5301 \\
 5 &    -28058.4828 &    -64106.0823 &    124464.2583 &    -29951.3912 &    306175.2572 &    -82073.3650 \\
 6 &     32216.1179 &     78849.9255 &   -160153.3914 &     18670.2471 &   -364490.5335 &    118456.4867 \\
 7 &    -21842.1492 &    -55788.5163 &    118449.6452 &     -4559.7279 &    256109.3681 &    -95983.8513 \\
 8 &      7927.7781 &     20463.3729 &    -45841.3613 &       -70.1676 &    -95377.6143 &     40024.2265 \\
 9 &      -966.9874 &     -1991.3079 &      5864.5150 &      -593.6420 &     11176.9161 &     -5694.0353 \\
10 &       -90.8323 &      -733.5016 &       492.0410 &       687.7279 &      1865.1290 &      -136.7890 \\
11 &       -82.7179 &        49.7887 &       627.7072 &      -267.6848 &       620.4204 &      -862.0290 \\
12 &        39.8748 &       -41.6052 &      -314.5589 &       144.9254 &      -281.0109 &       457.1299 \\
13 &        -1.5923 &        81.4695 &        33.7649 &      -103.1123 &      -107.7827 &      -106.6717 \\
14 &         5.9723 &       -17.3877 &       -49.2034 &        36.1700 &       -25.2992 &        78.3968 \\
\hline
\end{tabular}
}
\end{center}
\end{table}

\begin{table}
\caption{Constants $c_i(S,L,J;n)$ (in GeV) for
the wave functions of mesons with $L=1$}
\begin{center} {\scriptsize
\begin{tabular}{|r|r|r|r|r|r|r|}
\hline
& $a_0(1P)$ & $a_0(2P)$ & $a_0(3P)$ & $a_0(4P)$ & $a_0(5P)$ & $a_0(6P)$ \\
\hline
$i$ & $\psi_{1}^{(1,1,0)}$ & $\psi_{2}^{(1,1,0)}$ & $\psi_{3}^{(1,1,0)}$ & $\psi_{4}^{(1,1,0)}$  & $\psi_{5}^{(1,1,0)}$ & $\psi_{6}^{(1,1,0)}$  \\
\hline
 1 &        42.4371 &        79.7654 &       181.7393 &       552.3046 &     -1290.4221 &      1321.1855 \\
 2 &         8.2985 &       174.5586 &        52.8948 &     -3509.0430 &     13116.2453 &    -15761.5826 \\
 3 &      -119.5112 &     -1866.0998 &     -4767.6115 &      6343.4037 &    -50766.6682 &     71435.2573 \\
 4 &         9.0451 &      3990.4439 &     14898.6777 &       302.5701 &     98912.8327 &   -162408.9215 \\
 5 &       205.9024 &     -4036.6287 &    -19963.7100 &    -12748.5059 &   -105685.8815 &    201420.7033 \\
 6 &      -213.0162 &      2137.9247 &     13294.9892 &     14124.0468 &     60954.8263 &   -133873.7864 \\
 7 &        74.1245 &      -506.2077 &     -3795.7597 &     -5290.2962 &    -15642.5766 &     39165.6001 \\
 8 &        -0.0217 &         0.1123 &         0.5500 &         0.9631 &         3.0992 &        -7.4621 \\
 9 &        -0.9078 &         8.6218 &         6.4111 &       -28.7501 &       206.0889 &      -207.3682 \\
10 &        -2.2698 &         1.2428 &       131.1635 &       328.4528 &       181.1010 &     -1298.3571 \\
11 &         0.0222 &         2.9563 &       -20.0748 &       -66.3260 &        41.3323 &       157.1372 \\
12 &        -0.1182 &         0.3410 &         2.4538 &         8.2962 &         7.3024 &       -36.4266 \\
13 &         0.4600 &        -1.3504 &       -10.9087 &       -31.1803 &       -34.2322 &       142.7045 \\
14 &        -0.1234 &         0.1221 &         3.0370 &        11.2496 &         2.2632 &       -36.3022 \\
\hline
\hline
& $a_1(1P)$ & $a_1(2P)$ & $a_1(3P)$ & $a_1(4P)$ & $a_1(5P)$ & $a_1(6P)$ \\
\hline
$i$ & $\psi_{1}^{(1,1,1)}$ & $\psi_{2}^{(1,1,1)}$ & $\psi_{3}^{(1,1,1)}$ & $\psi_{4}^{(1,1,1)}$  & $\psi_{5}^{(1,1,1)}$ & $\psi_{6}^{(1,1,1)}$  \\
\hline
 1 &        34.2013 &       -52.2565 &        71.0776 &     -1306.0665 &      1501.5463 &      -457.8522 \\
 2 &        73.3267 &      -512.9001 &      1204.0408 &     12858.8666 &    -17980.8928 &      6811.9717 \\
 3 &      -618.2493 &      4043.3457 &    -10596.7170 &    -51117.9731 &     84263.2503 &    -37499.1623 \\
 4 &      1388.6534 &    -10764.8858 &     32068.1009 &    109346.2651 &   -206029.5860 &    103756.0404 \\
 5 &     -1689.1914 &     15097.5031 &    -49715.6179 &   -138966.7394 &    291090.4795 &   -161427.9679 \\
 6 &      1232.9066 &    -12248.3958 &     43471.4387 &    106983.3914 &   -243452.1939 &    145646.8677 \\
 7 &      -517.8134 &      5556.0836 &    -20871.7341 &    -47028.6214 &    114256.7199 &    -72620.6093 \\
 8 &        96.2280 &     -1092.2494 &      4300.7002 &      9062.0454 &    -23245.5299 &     15532.7919 \\
 9 &         4.1967 &       -48.5233 &       198.3314 &       389.4796 &     -1061.4740 &       755.4667 \\
10 &        -1.8198 &        17.2571 &       -84.5512 &      -141.4949 &       443.4312 &      -350.3904 \\
11 &        -0.0465 &         4.6670 &       -18.9098 &       -31.8081 &        99.1522 &       -77.0194 \\
12 &        -0.7730 &         9.1757 &       -35.9839 &       -75.0456 &       195.1395 &      -133.3575 \\
13 &         0.2384 &        -3.1342 &        13.5568 &        24.7992 &       -72.1201 &        54.0402 \\
14 &         0.0291 &        -0.5382 &         1.7905 &         4.2442 &        -9.9964 &         6.2249 \\
\hline
\hline
& $a_2(1P)$ & $a_2(2P)$ & $a_2(3P)$ & $a_2(4P)$ & $a_2(5P)$ & $a_2(6P)$ \\
\hline
$i$ & $\psi_{1}^{(1,1,2)}$ & $\psi_{2}^{(1,1,2)}$ & $\psi_{3}^{(1,1,2)}$ & $\psi_{4}^{(1,1,2)}$  & $\psi_{5}^{(1,1,2)}$ & $\psi_{6}^{(1,1,2)}$  \\
\hline
 1 &        32.2197 &       -77.8059 &      -210.1186 &       647.4775 &     -1283.4171 &     -1100.4179 \\
 2 &        20.0217 &      -166.6368 &       408.0845 &     -4983.7876 &     13525.0336 &     13244.2352 \\
 3 &      -216.6993 &      2089.0668 &      2776.4236 &     14397.9264 &    -55685.6519 &    -61558.2594 \\
 4 &       312.8691 &     -5329.3392 &    -11625.2497 &    -20482.8168 &    119270.9988 &    146729.0492 \\
 5 &      -175.0944 &      6698.4919 &     18318.0652 &     15619.1465 &   -146634.2681 &   -197055.7611 \\
 6 &         9.4278 &     -4684.8164 &    -14419.4714 &     -6876.1351 &    105022.3635 &    150004.9984 \\
 7 &        29.2027 &      1719.0994 &      5260.0087 &      2638.3256 &    -40469.9671 &    -58080.7068 \\
 8 &        -8.1701 &      -222.0968 &      -331.7885 &     -1248.4030 &      5911.0500 &      6575.8334 \\
 9 &         0.1830 &         5.3827 &        10.6671 &        19.1829 &      -120.3026 &      -143.1095 \\
10 &        -1.1814 &       -45.6707 &      -345.7367 &       472.1850 &       905.6950 &      2576.6751 \\
11 &         0.5240 &        22.7561 &       171.3334 &      -229.4744 &      -473.8713 &     -1319.8969 \\
12 &         0.1126 &        -0.8057 &       -11.5255 &        20.2313 &        12.2486 &        68.1575 \\
13 &        -0.0765 &        -1.1308 &        -6.0498 &         5.8793 &        26.0189 &        57.1325 \\
14 &         0.0119 &         0.0191 &        -0.3484 &         0.8055 &        -0.0988 &         2.1049 \\
\hline
\end{tabular}
}
\end{center}
\end{table}

\begin{table}
\caption{Constants $c_i(S,L,J;n)$ (in GeV) for
the wave functions of mesons with $L=1$}
\begin{center} {\scriptsize
\begin{tabular}{|r|r|r|r|r|r|r|}
\hline
& $b_1(1P)$ & $b_1(2P)$ & $b_1(3P)$ & $b_1(4P)$ & $b_1(5P)$ & $b_1(6P)$ \\
\hline
$i$ & $\psi_{1}^{(0,1,1)}$ & $\psi_{2}^{(0,1,1)}$ & $\psi_{3}^{(0,1,1)}$ & $\psi_{4}^{(0,1,1)}$  & $\psi_{5}^{(0,1,1)}$ & $\psi_{6}^{(0,1,1)}$  \\
\hline
 1 &        39.8521 &      -101.1249 &       289.7289 &      -676.1740 &       913.4728 &      -867.3699 \\
 2 &        27.9683 &        59.8776 &     -1349.0671 &      5529.2711 &     -9320.4160 &      9958.0589 \\
 3 &      -436.3532 &      1394.5931 &      1204.7709 &    -17483.5077 &     37240.3064 &    -44614.6046 \\
 4 &       963.5384 &     -4304.2438 &      3319.6286 &     28515.1031 &    -77936.1123 &    104095.0015 \\
 5 &     -1103.7396 &      5943.5023 &     -8934.1761 &    -26716.8499 &     95207.1565 &   -140358.8466 \\
 6 &       766.6038 &     -4579.2220 &      9021.3411 &     15134.0348 &    -70394.1107 &    112613.0501 \\
 7 &      -323.2662 &      1989.9446 &     -4456.2065 &     -5436.3308 &     30713.5583 &    -51719.4569 \\
 8 &        72.5557 &      -418.6745 &       922.3306 &      1319.3334 &     -6909.2056 &     11497.3761 \\
 9 &        -7.9236 &        32.9892 &       -72.1173 &      -124.7161 &       590.7466 &      -955.0237 \\
10 &         5.9239 &       -29.6551 &       128.3077 &      -100.6490 &      -307.4354 &       854.7824 \\
11 &        -3.9474 &        19.7152 &       -78.8994 &        45.7275 &       228.3113 &      -575.0888 \\
12 &        -0.0915 &        -0.0605 &         2.2162 &        -7.7219 &         9.0296 &        -2.3901 \\
13 &         0.8846 &        -3.3118 &         9.6400 &         5.8656 &       -54.4779 &       100.5059 \\
14 &        -0.1937 &         0.6423 &        -1.6426 &        -2.0444 &        11.6284 &       -19.5816 \\
\hline
\hline
& $f_2(1Pn\bar n)$ & $f_2(2Pn\bar n)$ & $f_2(3Pn\bar n)$ &
 $f_2(4Pn\bar n)$ & $f_2(5Pn\bar n)$ & $f_2(6Pn\bar n)$ \\
\hline
$i$ & $\psi_{1}^{(1,1,2)}$ & $\psi_{2}^{(1,1,2)}$ & $\psi_{3}^{(1,1,2)}$ & $\psi_{4}^{(1,1,2)}$  & $\psi_{5}^{(1,1,2)}$ & $\psi_{6}^{(1,1,2)}$  \\
\hline
 1 &        32.2197 &       -77.8059 &      -210.1186 &       647.4775 &     -1283.4171 &     -1100.4179 \\
 2 &        20.0217 &      -166.6368 &       408.0845 &     -4983.7876 &     13525.0336 &     13244.2352 \\
 3 &      -216.6993 &      2089.0668 &      2776.4236 &     14397.9264 &    -55685.6519 &    -61558.2594 \\
 4 &       312.8691 &     -5329.3392 &    -11625.2497 &    -20482.8168 &    119270.9988 &    146729.0492 \\
 5 &      -175.0944 &      6698.4919 &     18318.0652 &     15619.1465 &   -146634.2681 &   -197055.7611 \\
 6 &         9.4278 &     -4684.8164 &    -14419.4714 &     -6876.1351 &    105022.3635 &    150004.9984 \\
 7 &        29.2027 &      1719.0994 &      5260.0087 &      2638.3256 &    -40469.9671 &    -58080.7068 \\
 8 &        -8.1701 &      -222.0968 &      -331.7885 &     -1248.4030 &      5911.0500 &      6575.8334 \\
 9 &         0.1830 &         5.3827 &        10.6671 &        19.1829 &      -120.3026 &      -143.1095 \\
10 &        -1.1814 &       -45.6707 &      -345.7367 &       472.1850 &       905.6950 &      2576.6751 \\
11 &         0.5240 &        22.7561 &       171.3334 &      -229.4744 &      -473.8713 &     -1319.8969 \\
12 &         0.1126 &        -0.8057 &       -11.5255 &        20.2313 &        12.2486 &        68.1575 \\
13 &        -0.0765 &        -1.1308 &        -6.0498 &         5.8793 &        26.0189 &        57.1325 \\
14 &         0.0119 &         0.0191 &        -0.3484 &         0.8055 &        -0.0988 &         2.1049 \\
\hline
\hline
& $f_2(1Ps\bar s)$ & $f_2(2Ps\bar s)$ & $f_2(3Ps\bar s)$ &
 $f_2(4Ps\bar s)$ & $f_2(5Ps\bar s)$ & $f_2(6Ps\bar s)$ \\
\hline
$i$ & $\psi_{1}^{(1,1,2)}$ & $\psi_{2}^{(1,1,2)}$ & $\psi_{3}^{(1,1,2)}$ & $\psi_{4}^{(1,1,2)}$  & $\psi_{5}^{(1,1,2)}$ & $\psi_{6}^{(1,1,2)}$  \\
\hline
 1 &        31.7635 &        56.6130 &       116.4500 &      -335.2025 &      -895.9262 &      1168.5391 \\
 2 &        17.1760 &       173.3018 &       238.3886 &      1676.4843 &      8581.7170 &    -13412.9461 \\
 3 &      -198.4308 &     -1678.8511 &     -4390.6973 &      -769.7939 &    -31884.4208 &     59912.8528 \\
 4 &       253.7153 &      3908.3915 &     13430.5011 &     -8815.0581 &     61154.1382 &   -138251.2620 \\
 5 &       -82.1767 &     -4519.2423 &    -19130.0911 &     20834.4424 &    -67025.8946 &    181311.7783 \\
 6 &       -62.1099 &      2925.2396 &     14509.7708 &    -19967.1900 &     43185.3719 &   -136704.6221 \\
 7 &        50.6343 &     -1020.4285 &     -5488.1799 &      8382.9610 &    -15875.6151 &     54360.2519 \\
 8 &        -5.5760 &       147.7925 &       618.3152 &      -700.6968 &      2950.0850 &     -7789.1900 \\
 9 &         0.0151 &        -7.7254 &       -21.7562 &         7.8155 &      -135.8152 &       262.5597 \\
10 &        -4.3365 &        13.8302 &       222.9730 &      -583.6364 &       -93.1980 &     -1617.7140 \\
11 &         2.0929 &        -5.9821 &      -111.4150 &       297.6236 &        53.7367 &       834.0838 \\
12 &        -0.0292 &         0.2070 &         7.0651 &       -17.9093 &        -8.1068 &       -38.8087 \\
13 &        -0.1078 &         0.2639 &         3.6654 &       -10.2888 &         0.1797 &       -34.3046 \\
14 &         0.0015 &        -0.0397 &         0.3942 &        -1.3137 &        -1.1680 &        -2.4689 \\
\hline
\end{tabular}
}
\end{center}
\end{table}

\begin{table}
\caption{Constants $c_i(S,L,J;n)$ (in GeV) for
the wave functions of mesons with $L=2$}
\begin{center} {\scriptsize
\begin{tabular}{|r|r|r|r|r|r|r|}
\hline
& $\rho_1(1D)$ & $\rho_1(2D)$ & $\rho_1(3D)$ & $\rho_1(4D)$ & $\rho_1(5D)$ & $\rho_1(6D)$ \\
\hline
$i$ & $\psi_{1}^{(1,2,1)}$ & $\psi_{2}^{(1,2,1)}$ & $\psi_{3}^{(1,2,1)}$ & $\psi_{4}^{(1,2,1)}$  & $\psi_{5}^{(1,2,1)}$ & $\psi_{6}^{(1,2,1)}$  \\
\hline
 1 &        32.6658 &         1.9163 &       295.8938 &      1109.3066 &     -2972.4612 &      1576.8739 \\
 2 &      -297.9197 &       -20.8031 &     -2587.2263 &     -9686.9261 &     25655.3732 &    -13528.9738 \\
 3 &      1030.3273 &        85.0553 &      8635.8893 &     32404.0185 &    -84048.6359 &     44382.9538 \\
 4 &     -1720.3250 &      -207.3837 &    -13721.7857 &    -52043.5015 &    130987.5209 &    -70210.9080 \\
 5 &      1257.2172 &       242.8759 &      9530.7617 &     36934.5548 &    -90093.3664 &     49448.7071 \\
 6 &        68.1505 &         4.0628 &       206.3101 &      1219.6605 &     -2593.2892 &      1668.7638 \\
 7 &      -702.1197 &      -203.4247 &     -4305.9309 &    -18749.1779 &     42830.4796 &    -25346.4851 \\
 8 &       419.2221 &       125.4426 &      2314.3640 &     10789.0751 &    -24056.1865 &     14784.8329 \\
 9 &      -113.3895 &       -25.0314 &      -521.0553 &     -2650.0621 &      5810.3996 &     -3683.5040 \\
10 &        68.2765 &        16.0844 &       378.0685 &      1715.0504 &     -3891.1978 &      2329.7965 \\
11 &       -58.4977 &       -16.6735 &      -340.7538 &     -1533.5489 &      3471.2170 &     -2088.1528 \\
12 &        19.6361 &         5.5328 &       110.3439 &       508.2071 &     -1142.4633 &       696.4236 \\
13 &         0.2485 &         0.0411 &         1.0932 &         5.2416 &       -11.7110 &         6.9974 \\
14 &        -1.1270 &        -0.2443 &        -5.1114 &       -25.8453 &        56.5560 &       -35.6361 \\
\hline
\hline
& $\rho_3(1D)$ & $\rho_3(2D)$ & $\rho_3(3D)$ & $\rho_3(4D)$ & $\rho_3(5D)$ & $\rho_3(6D)$ \\
\hline
$i$ & $\psi_{1}^{(1,2,3)}$ & $\psi_{2}^{(1,2,3)}$ & $\psi_{3}^{(1,2,3)}$ & $\psi_{4}^{(1,2,3)}$  & $\psi_{5}^{(1,2,3)}$ & $\psi_{6}^{(1,2,3)}$  \\
\hline
 1 &         2.7159 &         0.2776 &       -35.8536 &       -51.1641 &      -244.5533 &       377.8216 \\
 2 &       -28.9187 &        11.5948 &       345.1902 &       678.5250 &      2151.6525 &     -4485.1440 \\
 3 &       114.9817 &      -100.7846 &     -1263.4123 &     -3288.1168 &     -6799.5737 &     19118.2096 \\
 4 &      -228.6952 &       325.0888 &      2187.8281 &      7495.6151 &      9688.7068 &    -38191.8418 \\
 5 &       228.9398 &      -475.1771 &     -1814.1391 &     -8396.0297 &     -6454.1426 &     38608.9724 \\
 6 &      -101.5343 &       282.1589 &       660.9834 &      4254.9784 &      1831.5183 &    -18287.5625 \\
 7 &        14.1041 &       -36.5480 &       -84.3804 &      -566.5141 &      -308.2865 &      2423.3962 \\
 8 &        -2.7182 &         6.3595 &        15.4417 &       111.3330 &        43.9165 &      -501.4362 \\
 9 &         5.0150 &       -34.1846 &         5.1873 &      -431.2745 &       239.5250 &      1699.3325 \\
10 &        -2.0534 &        17.1290 &        -9.7324 &       213.2919 &      -158.1876 &      -830.8559 \\
11 &         0.0405 &        -0.0746 &        -1.4433 &        -0.6579 &        -6.2614 &         5.2875 \\
12 &         0.1601 &        -1.2656 &         1.8759 &       -16.8188 &        18.0360 &        64.8080 \\
13 &        -0.0278 &         0.0356 &         0.1298 &         0.7826 &         0.4594 &        -3.4686 \\
14 &        -0.0107 &         0.0212 &         0.0501 &         0.3805 &         0.1459 &        -1.6595 \\
\hline
\hline
& $\pi_2(1D)$ & $\pi_2(2D)$ & $\pi_2(3D)$ & $\pi_2(4D)$ & $\pi_2(5D)$ & $\pi_2(6D)$ \\
\hline
$i$ & $\psi_{1}^{(0,2,2)}$ & $\psi_{2}^{(0,2,2)}$ & $\psi_{3}^{(0,2,2)}$ & $\psi_{4}^{(0,2,2)}$  & $\psi_{5}^{(0,2,2)}$ & $\psi_{6}^{(0,2,2)}$  \\
\hline
 1 &         1.7299 &        -4.5924 &       -30.4454 &       -25.0916 &       236.8928 &       -10.9398 \\
 2 &       -21.7161 &        31.2366 &       317.1714 &        47.5520 &     -2266.4156 &      -960.9940 \\
 3 &        95.0803 &       -63.9765 &     -1269.6697 &       641.8071 &      8037.3068 &      7243.4624 \\
 4 &      -209.5669 &       -12.3479 &      2466.5019 &     -2875.2814 &    -13493.1939 &    -18963.7627 \\
 5 &       232.2060 &       190.4915 &     -2406.5203 &      4478.4005 &     11606.9964 &     23306.1246 \\
 6 &      -116.6391 &      -192.6014 &      1140.4369 &     -3020.3383 &     -5014.5881 &    -13827.9449 \\
 7 &        21.3512 &        53.9102 &      -224.0370 &       738.5291 &       984.2641 &      3211.6746 \\
 8 &        -2.8057 &        -4.0451 &        28.6196 &       -70.9307 &      -131.0914 &      -342.1592 \\
 9 &         3.6836 &        15.0146 &       -23.3009 &       178.3822 &        26.5892 &       709.9538 \\
10 &        -1.5247 &        -9.1777 &         8.6870 &      -105.1379 &         8.3760 &      -409.4915 \\
11 &         0.2238 &         0.0350 &        -1.8229 &         1.7643 &         8.5490 &        10.7963 \\
12 &        -0.0083 &         0.8066 &         0.2574 &         8.8653 &        -5.7694 &        33.1803 \\
13 &        -0.0138 &        -0.0196 &         0.1708 &        -0.4001 &        -0.7849 &        -1.9340 \\
14 &        -0.0068 &        -0.0125 &         0.0775 &        -0.2125 &        -0.3463 &        -0.9910 \\
\hline
\end{tabular}
}
\end{center}
\end{table}

\begin{table}
\caption{Constants $c_i(S,L,J;n)$ (in GeV) for
the wave functions of mesons with $L=2$ }
\begin{center} {\scriptsize
\begin{tabular}{|r|r|r|r|r|r|r|}
\hline
& $\phi_3(1D)$ & $\phi_3(2D)$ & $\phi_3(3D)$ & $\phi_3(4D)$ & $\phi_3(5D)$ & $\phi_3(6D)$ \\
\hline
$i$ & $\psi_{1}^{(1,2,3)}$ & $\psi_{2}^{(1,2,3)}$ & $\psi_{3}^{(1,2,3)}$ & $\psi_{4}^{(1,2,3)}$  & $\psi_{5}^{(1,2,3)}$ & $\psi_{6}^{(1,2,3)}$  \\
\hline
 1 &        12.7970 &       -16.5406 &       131.3023 &       175.9948 &      1159.9649 &      2336.2002 \\
 2 &      -132.9591 &       131.2478 &     -1400.8339 &     -1057.6352 &    -11576.8783 &    -17282.6320 \\
 3 &       545.7787 &      -328.1630 &      5638.4339 &      1282.1727 &     43302.9673 &     50238.5349 \\
 4 &     -1133.9123 &       198.2831 &    -11023.2534 &      2464.8624 &    -80609.6554 &    -75397.6670 \\
 5 &      1257.8987 &       266.6776 &     11305.5592 &     -6986.7989 &     81345.9382 &     63000.7836 \\
 6 &      -728.1703 &      -357.4173 &     -5962.3588 &      5787.4864 &    -43383.6503 &    -28494.7375 \\
 7 &       185.9914 &       103.1732 &      1310.5126 &     -1638.9434 &      9822.7300 &      5709.1796 \\
 8 &        -5.8150 &        -3.0432 &       -34.9643 &        59.9959 &      -302.0414 &      -164.8337 \\
 9 &         3.9102 &        26.5889 &        78.4045 &      -243.7804 &       706.5301 &       165.6968 \\
10 &        -4.7411 &       -16.9989 &       -53.2190 &       170.8107 &      -492.3601 &      -111.5096 \\
11 &        -0.9575 &        -0.4198 &        -3.1963 &         9.5569 &       -31.2590 &        -9.0747 \\
12 &         1.0046 &         1.5589 &         5.3910 &       -20.3250 &        55.9957 &        10.7975 \\
13 &        -0.0395 &         0.0358 &        -0.0643 &        -0.0061 &        -0.6678 &        -0.6516 \\
14 &        -0.0077 &         0.0216 &        -0.0215 &        -0.1466 &        -0.0529 &        -0.3559 \\
\hline
\end{tabular}
}
\end{center}
\end{table}

\begin{table}
\caption{Constants $c_i(S,L,J;n)$ (in GeV) for
the wave functions of mesons with $L=3$ }
\begin{center} {\scriptsize
\begin{tabular}{|r|r|r|r|r|r|r|}
\hline
& $a_2(1F)$ & $a_2(2F)$ & $a_2(3F)$ & $a_2(4F)$ & $a_2(5F)$ & $a_2(6F)$ \\
\hline
$i$ & $\psi_{1}^{(1,3,2)}$ & $\psi_{2}^{(1,3,2)}$ & $\psi_{3}^{(1,3,2)}$ & $\psi_{4}^{(1,3,2)}$  & $\psi_{5}^{(1,3,2)}$ & $\psi_{6}^{(1,3,2)}$  \\
\hline
 1 &       302.5298 &      3108.5771 &     -4814.1850 &     -3261.0719 &      2682.3682 &     -2424.4730 \\
 2 &      -143.3529 &    -16363.8999 &     29608.0652 &     22339.5339 &    -19180.4679 &     17736.4524 \\
 3 &     -1820.4311 &     33890.9895 &    -70605.8774 &    -58593.2358 &     52807.6894 &    -50085.1286 \\
 4 &      3544.0813 &    -34294.0253 &     81404.4846 &     73808.0123 &    -70047.4087 &     68368.7967 \\
 5 &     -2486.5011 &     15588.3523 &    -42971.6679 &    -42810.1076 &     43057.3359 &    -43360.3298 \\
 6 &       505.7921 &      -751.9773 &      4532.7093 &      5800.9589 &     -6528.4324 &      6781.1326 \\
 7 &        33.3061 &      -272.7190 &       747.1767 &       788.7331 &      -871.6059 &       969.3650 \\
 8 &       224.4548 &     -2230.2392 &      5730.9132 &      5836.8855 &     -6325.0912 &      7073.3615 \\
 9 &      -222.8275 &      1735.8163 &     -4925.5549 &     -5402.9935 &      6233.6463 &     -7291.5857 \\
10 &        66.4973 &      -422.2953 &      1339.5158 &      1592.2922 &     -1974.9107 &      2446.8811 \\
11 &        -3.4375 &         1.5454 &       -40.8223 &       -76.1636 &       122.8760 &      -178.2320 \\
12 &        -0.4697 &         4.0630 &       -12.2221 &       -14.6895 &        19.7422 &       -27.3408 \\
13 &         0.1083 &        -1.0426 &         2.5864 &         2.4508 &        -2.3343 &         2.1666 \\
14 &        -0.3887 &         3.4138 &        -8.9913 &        -9.1924 &         9.6794 &       -10.1976 \\
\hline
\hline
& $a_3(1F)$ & $a_3(2F)$ & $a_3(3F)$ & $a_3(4F)$ & $a_3(5F)$ & $a_3(6F)$ \\
\hline
$i$ & $\psi_{1}^{(1,3,3)}$ & $\psi_{2}^{(1,3,3)}$ & $\psi_{3}^{(1,3,3)}$ & $\psi_{4}^{(1,3,3)}$  & $\psi_{5}^{(1,3,3)}$ & $\psi_{6}^{(1,3,3)}$  \\
\hline
 1 &      -185.6131 &     -1273.6139 &     -2824.8371 &     -3304.5424 &      3223.3854 &      3056.8805 \\
 2 &       100.1634 &      5502.8339 &     15900.4766 &     21342.1692 &    -22465.1172 &    -22346.9515 \\
 3 &       997.5452 &     -8945.0906 &    -35307.5312 &    -54519.8991 &     62298.6654 &     65295.7024 \\
 4 &     -2016.2799 &      6492.3037 &     39328.0259 &     70562.2824 &    -88115.5292 &    -97910.4387 \\
 5 &      1587.7723 &     -1474.4759 &    -22363.9804 &    -47827.8049 &     65991.6227 &     78406.5798 \\
 6 &      -509.7646 &      -457.3439 &      5241.1083 &     14381.9488 &    -22431.8112 &    -28904.4118 \\
 7 &         0.9615 &        -5.2535 &       -29.5094 &       -53.1605 &        59.1638 &        38.2607 \\
 8 &        17.7201 &       232.8948 &       263.4910 &      -357.5918 &      1384.8933 &      2617.0132 \\
 9 &         5.6231 &       -63.5036 &      -203.8880 &      -212.9591 &        48.0912 &      -260.1436 \\
10 &         3.8781 &         7.8110 &       -31.6415 &      -107.2999 &       179.8714 &       238.9973 \\
11 &        -3.6420 &       -20.5693 &         3.9441 &        94.5531 &      -203.4926 &      -302.4885 \\
12 &        -0.3596 &        10.8007 &        26.7768 &        19.6812 &         7.3775 &        43.2610 \\
13 &         0.6972 &        -0.4195 &        -9.8214 &       -22.3726 &        32.0206 &        38.4703 \\
14 &        -0.1490 &        -0.8245 &         0.2198 &         4.1481 &        -9.2541 &       -14.6575 \\
\hline
\hline
& $b_3(1F)$ & $b_3(2F)$ & $b_3(3F)$ & $b_3(4F)$ & $b_3(5F)$ & $b_3(6F)$ \\
\hline
$i$ & $\psi_{1}^{(0,3,3)}$ & $\psi_{2}^{(0,3,3)}$ & $\psi_{3}^{(0,3,3)}$ & $\psi_{4}^{(0,3,3)}$  & $\psi_{5}^{(0,3,3)}$ & $\psi_{6}^{(0,3,3)}$  \\
\hline
 1 &       -42.3814 &      -688.7727 &      4871.3033 &     -6800.6553 &      3306.7862 &      2167.3909 \\
 2 &      -700.1877 &      1416.6923 &    -30922.8704 &     49960.3764 &    -26808.6339 &    -18153.6700 \\
 3 &      2996.2839 &      2579.9739 &     80377.0755 &   -148188.4558 &     86821.4959 &     60811.3974 \\
 4 &     -4886.9006 &    -10605.0277 &   -110657.5998 &    229300.7333 &   -145016.6326 &   -105049.8860 \\
 5 &      4029.3580 &     12572.9407 &     84979.6664 &   -194602.3500 &    131244.3438 &     98109.2312 \\
 6 &     -1569.0269 &     -6072.3884 &    -32063.7769 &     79614.6525 &    -56506.2810 &    -43390.4267 \\
 7 &        -0.6830 &        41.7580 &       -23.3734 &      -208.2752 &       306.6619 &       354.9735 \\
 8 &       255.5766 &      1070.7181 &      5175.5602 &    -13213.8899 &      9433.2311 &      7087.6133 \\
 9 &      -100.3425 &      -333.4997 &     -2066.1119 &      4736.4035 &     -3054.2265 &     -2016.5817 \\
10 &        18.4890 &        56.1607 &       381.9187 &      -847.4901 &       531.7885 &       343.3677 \\
11 &        -2.9991 &       -54.2606 &       -45.2623 &       373.8056 &      -423.9163 &      -451.1881 \\
12 &         0.0879 &        22.0627 &        -7.0167 &      -115.1478 &       161.4084 &       186.4986 \\
13 &         0.3332 &         1.9883 &         7.2523 &       -22.8240 &        20.0422 &        19.2270 \\
14 &        -0.0731 &        -2.7700 &        -0.7800 &        17.3809 &       -22.0744 &       -25.0482 \\
\hline
\end{tabular}
}
\end{center}
\end{table}

\begin{table}
\caption{Constants $c_i(S,L,J;n)$ (in GeV) for
the wave functions of mesons with $L=3$ }
\begin{center} {\scriptsize
\begin{tabular}{|r|r|r|r|r|r|r|}
\hline
& $a_4(1F)$ & $a_4(2F)$ & $a_4(3F)$ & $a_4(4F)$ & $a_4(5F)$ & $a_4(6F)$ \\
\hline
$i$ & $\psi_{1}^{(1,3,4)}$ & $\psi_{2}^{(1,3,4)}$ & $\psi_{3}^{(1,3,4)}$ & $\psi_{4}^{(1,3,4)}$  & $\psi_{5}^{(1,3,4)}$ & $\psi_{6}^{(1,3,4)}$  \\
\hline
 1 &        61.3146 &      -279.4264 &     -6335.7399 &      3793.3532 &     -2349.3250 &      1770.7208 \\
 2 &      -805.5228 &      -494.1693 &     38516.0215 &    -26710.9675 &     17820.9665 &    -13733.8853 \\
 3 &      2125.7747 &      4617.2211 &    -92101.7641 &     71500.9282 &    -51068.4486 &     40476.0076 \\
 4 &     -2401.6045 &     -7997.8369 &    107642.9762 &    -91068.5254 &     69089.8839 &    -56455.3468 \\
 5 &      1213.7027 &      5255.0393 &    -58196.3762 &     52536.8768 &    -41934.5759 &     35302.0531 \\
 6 &      -155.4700 &      -853.4752 &      8074.5612 &     -7758.0297 &      6510.9466 &     -5710.4149 \\
 7 &        54.5187 &       438.3874 &     -3773.4778 &      4071.3367 &     -4027.0585 &      4097.9035 \\
 8 &      -214.3060 &     -1528.9114 &     13678.2370 &    -14276.3588 &     13540.7770 &    -13205.8763 \\
 9 &       136.2202 &      1018.8150 &     -8962.3961 &      9476.4891 &     -9126.6891 &      9030.2398 \\
10 &        -1.9588 &       -13.6745 &       127.4084 &      -133.2553 &       129.7353 &      -130.8037 \\
11 &       -29.9559 &      -242.3946 &      2084.9041 &     -2255.0201 &      2241.8432 &     -2297.8267 \\
12 &        12.5501 &       100.9776 &      -872.4258 &       943.2049 &      -940.5795 &       970.7261 \\
13 &        -1.6887 &       -11.5254 &       104.7530 &      -108.0387 &       101.2025 &       -98.0783 \\
14 &        -0.0870 &        -1.6708 &        12.2570 &       -15.8234 &        19.4062 &       -24.0699 \\
\hline
\hline
& $f_2(1Fn\bar n)$ & $f_2(2Fn\bar n)$ & $f_2(3Fn\bar n)$ &
 $f_2(4Fn\bar n)$ & $f_2(5Fn\bar n)$ & $f_2(6Fn\bar n)$ \\
\hline
$i$ & $\psi_{1}^{(1,3,2)}$ & $\psi_{2}^{(1,3,2)}$ & $\psi_{3}^{(1,3,2)}$ & $\psi_{4}^{(1,3,2)}$  & $\psi_{5}^{(1,3,2)}$ & $\psi_{6}^{(1,3,2)}$  \\
\hline
 1 &       302.5298 &      3108.5771 &     -4814.1850 &     -3261.0719 &      2682.3682 &     -2424.4730 \\
 2 &      -143.3529 &    -16363.8999 &     29608.0652 &     22339.5339 &    -19180.4679 &     17736.4524 \\
 3 &     -1820.4311 &     33890.9895 &    -70605.8774 &    -58593.2358 &     52807.6894 &    -50085.1286 \\
 4 &      3544.0813 &    -34294.0253 &     81404.4846 &     73808.0123 &    -70047.4087 &     68368.7967 \\
 5 &     -2486.5011 &     15588.3523 &    -42971.6679 &    -42810.1076 &     43057.3359 &    -43360.3298 \\
 6 &       505.7921 &      -751.9773 &      4532.7093 &      5800.9589 &     -6528.4324 &      6781.1326 \\
 7 &        33.3061 &      -272.7190 &       747.1767 &       788.7331 &      -871.6059 &       969.3650 \\
 8 &       224.4548 &     -2230.2392 &      5730.9132 &      5836.8855 &     -6325.0912 &      7073.3615 \\
 9 &      -222.8275 &      1735.8163 &     -4925.5549 &     -5402.9935 &      6233.6463 &     -7291.5857 \\
10 &        66.4973 &      -422.2953 &      1339.5158 &      1592.2922 &     -1974.9107 &      2446.8811 \\
11 &        -3.4375 &         1.5454 &       -40.8223 &       -76.1636 &       122.8760 &      -178.2320 \\
12 &        -0.4697 &         4.0630 &       -12.2221 &       -14.6895 &        19.7422 &       -27.3408 \\
13 &         0.1083 &        -1.0426 &         2.5864 &         2.4508 &        -2.3343 &         2.1666 \\
14 &        -0.3887 &         3.4138 &        -8.9913 &        -9.1924 &         9.6794 &       -10.1976 \\
\hline
\hline
& $f_2(1Fs\bar s)$ & $f_2(2Fs\bar s)$ & $f_2(3Fs\bar s)$ &
 $f_2(4Fs\bar s)$ & $f_2(5Fs\bar s)$ & $f_2(6Fs\bar s)$ \\
\hline
$i$ & $\psi_{1}^{(1,3,2)}$ & $\psi_{2}^{(1,3,2)}$ & $\psi_{3}^{(1,3,2)}$ & $\psi_{4}^{(1,3,2)}$  & $\psi_{5}^{(1,3,2)}$ & $\psi_{6}^{(1,3,2)}$  \\
\hline
 1 &       231.0455 &      1397.2099 &     -2195.7858 &     -1633.9450 &      1101.4783 &      -757.7297 \\
 2 &      -297.0218 &     -6052.7648 &     11090.2912 &      9034.3538 &     -6273.3803 &      4353.3168 \\
 3 &      -340.4447 &      9852.3998 &    -20719.8583 &    -18146.2600 &     12864.5749 &     -8795.4222 \\
 4 &       681.3379 &     -7309.3654 &     16921.6464 &     15429.5960 &    -10646.2710 &      6524.4803 \\
 5 &      -222.4078 &      2015.3361 &     -4325.3140 &     -3335.3249 &      1116.1010 &      1011.8140 \\
 6 &       -94.8852 &       194.1384 &     -1321.2978 &     -2165.9299 &      2797.5823 &     -3405.6656 \\
 7 &        23.9313 &        30.9231 &       140.3553 &       315.6621 &      -422.2371 &       470.9011 \\
 8 &        21.5889 &      -171.7440 &       522.4358 &       711.7560 &      -836.3674 &      1028.8365 \\
 9 &        -3.3109 &        11.4442 &      -100.0134 &      -236.2070 &       386.1271 &      -570.1726 \\
10 &         4.9128 &        35.3511 &         1.5665 &        88.6043 &      -174.0941 &       242.8960 \\
11 &        -5.9620 &        -9.6868 &       -35.7124 &       -84.1853 &       116.6997 &      -131.8412 \\
12 &         1.2071 &        -0.2630 &        10.4401 &        19.1194 &       -24.4692 &        27.7934 \\
13 &         0.0594 &         0.0013 &         0.4416 &         0.7480 &        -0.8273 &         0.7159 \\
14 &         0.1767 &         0.1161 &         1.2881 &         2.4395 &        -3.0027 &         2.9259 \\
\hline
\end{tabular}
}
\end{center}
\end{table}

\begin{table}
\caption{Constants $c_i(S,L,J;n)$ (in GeV) for
the wave functions of mesons with $L=0$ }
\begin{center} {\scriptsize
\begin{tabular}{|r|r|r|r|r|r|r|}
\hline
& $\rho_3(1G)$ & $\rho_3(2G)$ & $\rho_3(3G)$ & $\rho_3(4G)$ & $\rho_3(5G)$ & $\rho_3(6G)$ \\
\hline
$i$ & $\psi_{1}^{(1,4,3)}$ & $\psi_{2}^{(1,4,3)}$ & $\psi_{3}^{(1,4,3)}$ & $\psi_{4}^{(1,4,3)}$  & $\psi_{5}^{(1,4,3)}$ & $\psi_{6}^{(1,4,3)}$  \\
\hline
 1 &      1154.9898 &      3997.7431 &     -4764.7329 &      3814.5732 &      3126.5090 &     -2727.8949 \\
 2 &     -3877.3074 &    -18832.5547 &     25685.2973 &    -22387.3916 &    -19219.6567 &     17208.4501 \\
 3 &      4872.7948 &     35104.5014 &    -54588.3818 &     51784.2762 &     46707.7823 &    -43054.1333 \\
 4 &     -2528.4558 &    -32539.5744 &     57587.7248 &    -59377.2151 &    -56362.6364 &     53619.0971 \\
 5 &       238.3859 &     14433.9604 &    -29079.6778 &     32454.3416 &     32388.1142 &    -31820.2906 \\
 6 &        -3.5692 &     -1161.5714 &      2412.5622 &     -2744.2147 &     -2774.6470 &      2747.8165 \\
 7 &       295.5542 &     -1490.0259 &      4400.1757 &     -5896.8647 &     -6579.2194 &      6967.7046 \\
 8 &      -161.8464 &       465.8255 &     -1686.9511 &      2420.9476 &      2804.7336 &     -3046.9726 \\
 9 &        -1.5301 &        99.9425 &      -216.7204 &       255.2614 &       270.4356 &      -281.4499 \\
10 &        13.7376 &      -129.0694 &       330.8061 &      -421.8631 &      -463.7937 &       491.4876 \\
11 &         0.3215 &        46.1738 &       -94.7864 &       106.9348 &       108.0270 &      -107.2939 \\
12 &         0.2398 &        -0.5283 &         2.1508 &        -3.1713 &        -3.7932 &         4.2158 \\
13 &        -0.9517 &        -3.8284 &         3.7693 &        -1.2982 &         0.9905 &        -2.8854 \\
14 &         0.2239 &         0.7053 &        -0.4845 &        -0.1453 &        -0.6845 &         1.1248 \\
\hline
\hline
& $\pi_4(1G)$ & $\pi_4(2G)$ & $\pi_4(3G)$ & $\pi_4(4G)$ & $\pi_4(5G)$ & $\pi_4(6G)$ \\
\hline
$i$ & $\psi_{1}^{(0,4,4)}$ & $\psi_{2}^{(0,4,4)}$ & $\psi_{3}^{(0,4,4)}$ & $\psi_{4}^{(0,4,4)}$  & $\psi_{5}^{(0,4,4)}$ & $\psi_{6}^{(0,4,4)}$  \\
\hline
 1 &      -910.2682 &     -2285.4269 &     -2544.6103 &      2193.2433 &     -1834.5475 &     -1542.4854 \\
 2 &      3296.7238 &     10036.7420 &     12377.7784 &    -11355.8564 &      9856.6894 &      8482.1168 \\
 3 &     -4826.6605 &    -17234.9296 &    -23262.5593 &     22679.8533 &    -20476.7609 &    -18088.5381 \\
 4 &      3506.8263 &     14308.2652 &     20849.2077 &    -21526.7049 &     20247.8664 &     18414.1598 \\
 5 &     -1120.1293 &     -5094.6942 &     -7903.5584 &      8590.2159 &     -8412.6647 &     -7884.9308 \\
 6 &       -85.4563 &      -442.1406 &      -737.0971 &       859.7994 &      -909.8070 &      -931.0430 \\
 7 &       197.0215 &      1049.2088 &      1790.4335 &     -2138.3661 &      2305.7629 &      2381.2048 \\
 8 &       -73.7778 &      -406.7605 &      -710.2853 &       871.4579 &      -970.3193 &     -1039.0872 \\
 9 &        11.0712 &        67.5285 &       122.0340 &      -152.4672 &       171.2164 &       184.0782 \\
10 &         4.6342 &        27.0370 &        48.2260 &       -60.2949 &        68.6923 &        75.8586 \\
11 &        -2.5520 &       -18.3205 &       -34.9905 &        45.4192 &       -52.4998 &       -57.7315 \\
12 &        -0.4969 &        -2.6183 &        -4.3866 &         5.1012 &        -5.3251 &        -5.3096 \\
13 &         0.5801 &         4.2384 &         8.1355 &       -10.5729 &        12.1852 &        13.2896 \\
14 &        -0.1078 &        -0.8541 &        -1.6837 &         2.2329 &        -2.6183 &        -2.8991 \\
\hline
\end{tabular}
}
\end{center}
\end{table}

\begin{figure}
\centerline{\epsfig{file=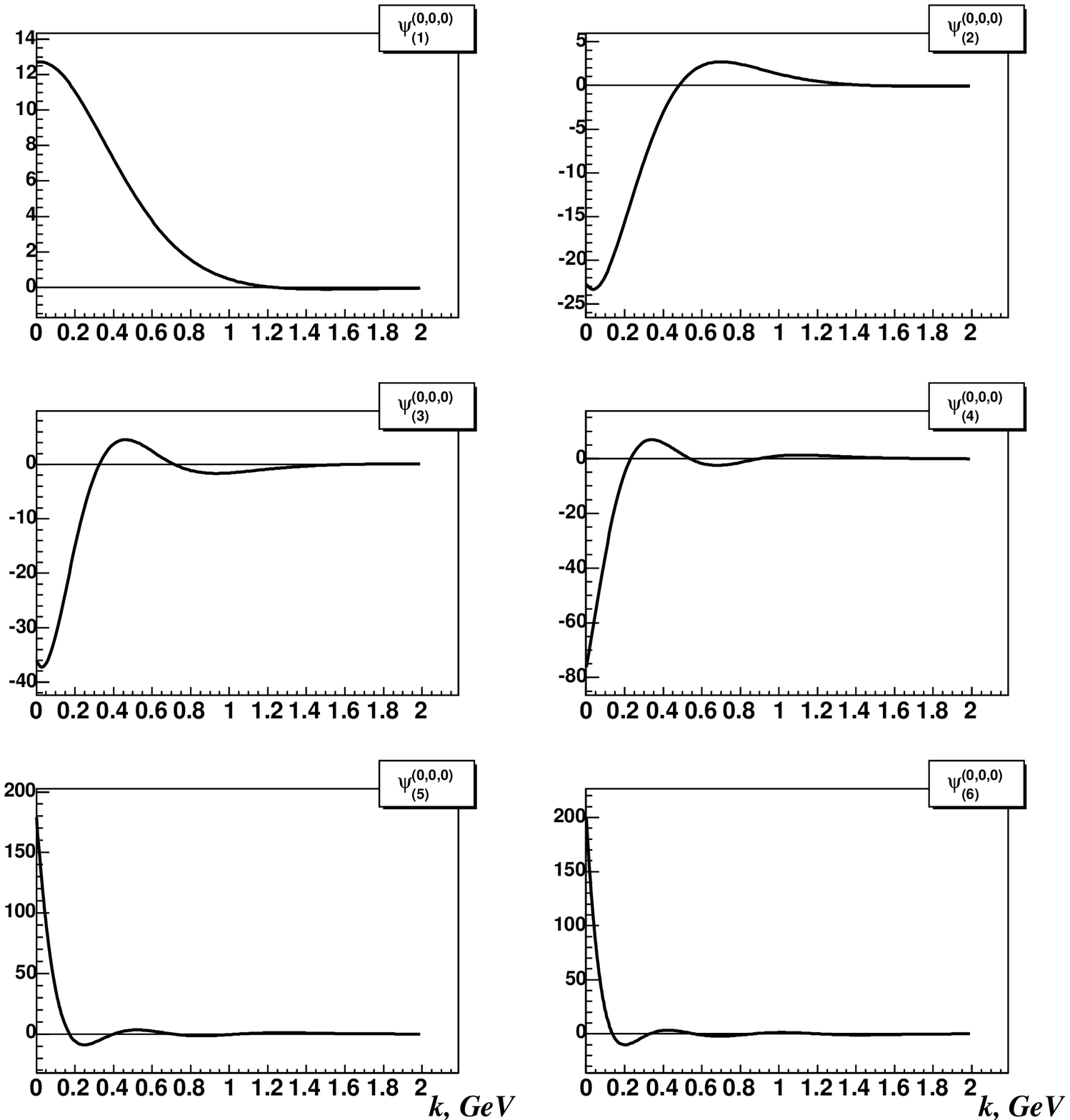,width=15cm}}
\caption{Wave functions of the  L=0 group
         ($\pi$-mesons with $n=1,2,3,4,5,6$).}
\end{figure}

\begin{figure}
\centerline{\epsfig{file=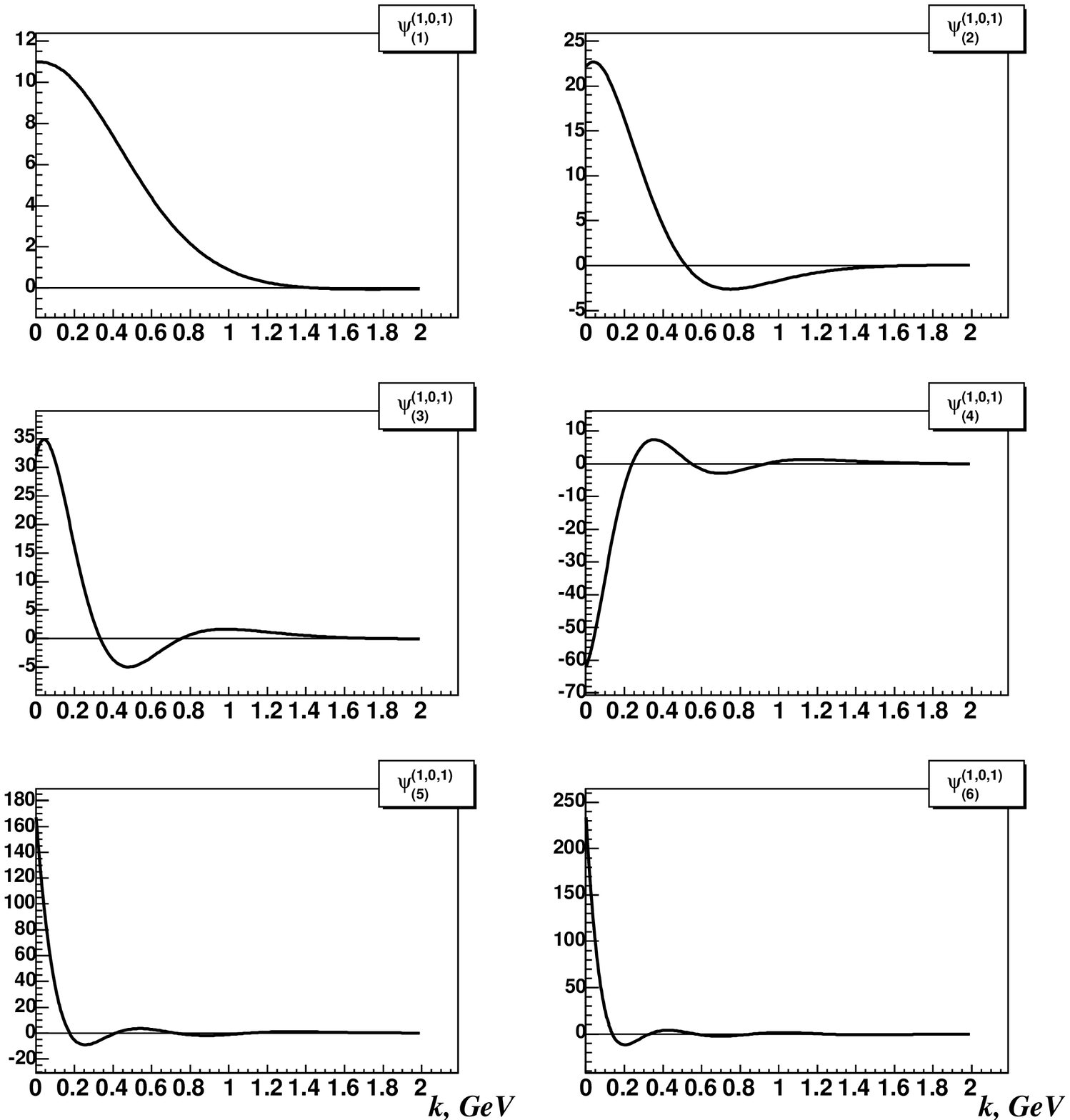,width=15cm}}
\caption{Wave functions of the  L=0 group
($\rho$- and $\omega$-mesons with $n=1,2,3,4,5,6$).}
\end{figure}

\begin{figure}
\centerline{\epsfig{file=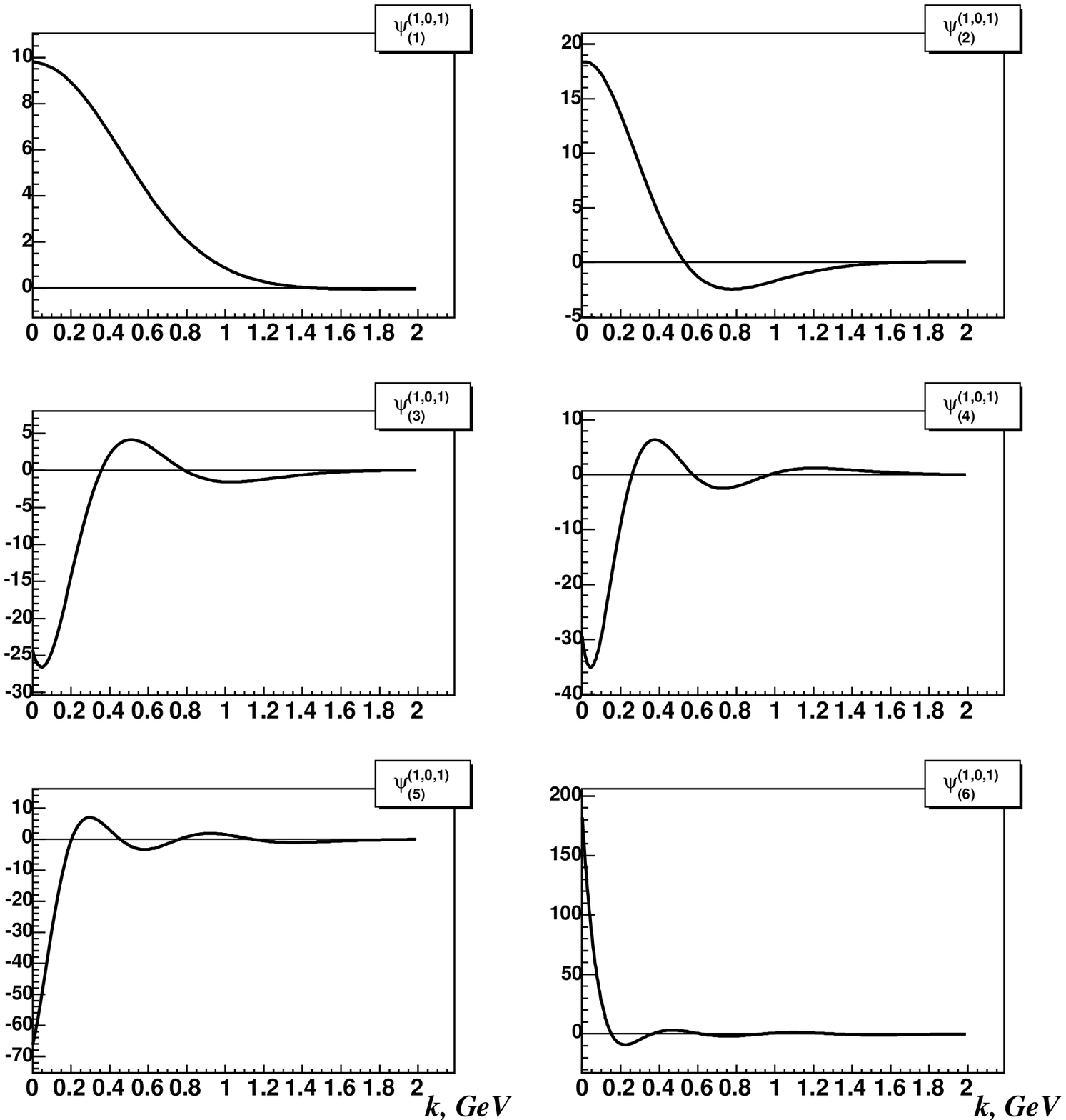,width=15cm}}
\caption{Wave functions of the  L=0 group
         ($\phi$-mesons with $n=1,2,3,4,5,6$).}
\end{figure}

\begin{figure}
\centerline{\epsfig{file=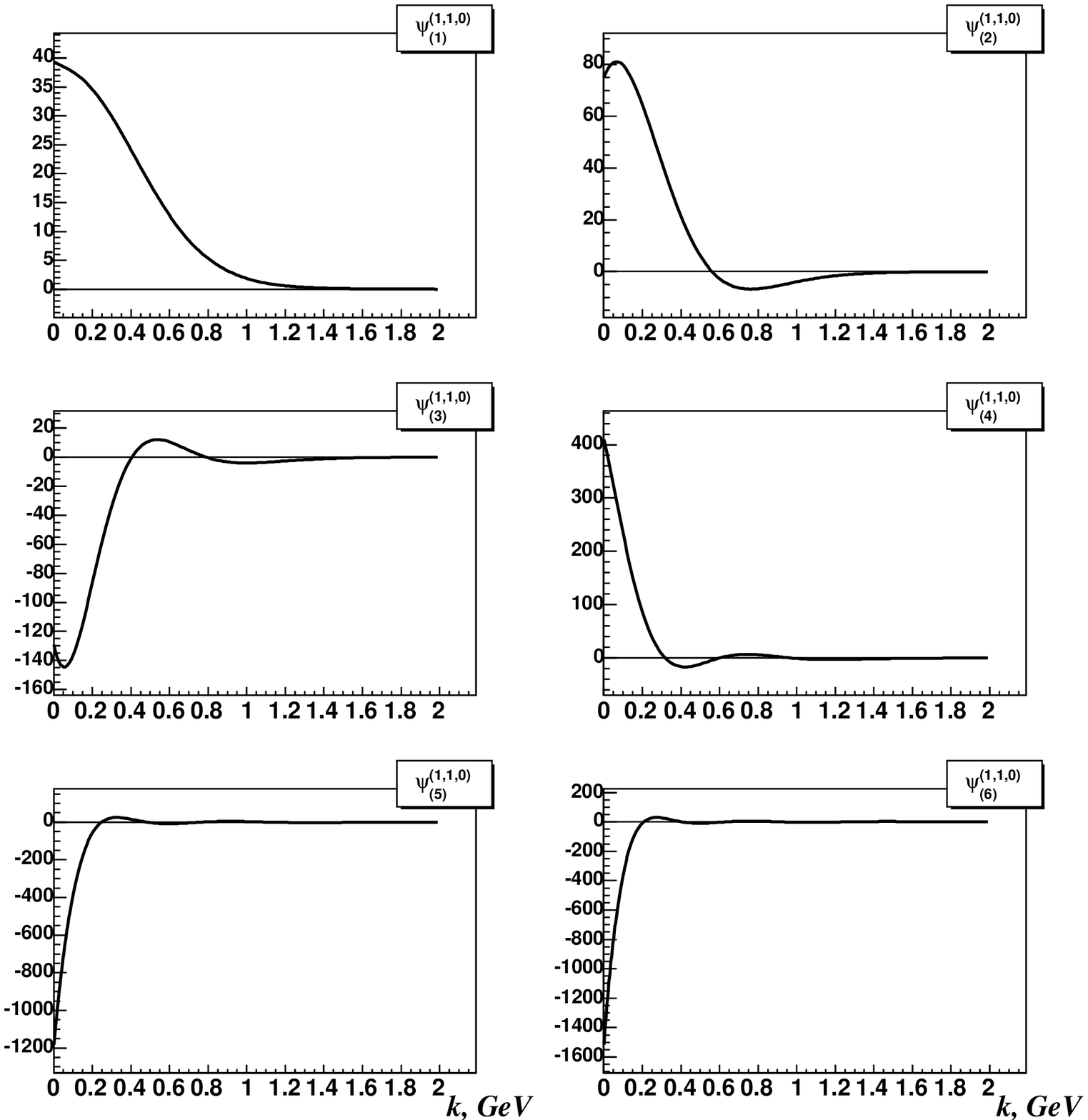,width=15cm}}
\caption{Wave functions of the  L=1 group
         ($a_0$-mesons with $n=1,2,3,4,5,6$).}
\end{figure}

\begin{figure}
\centerline{\epsfig{file=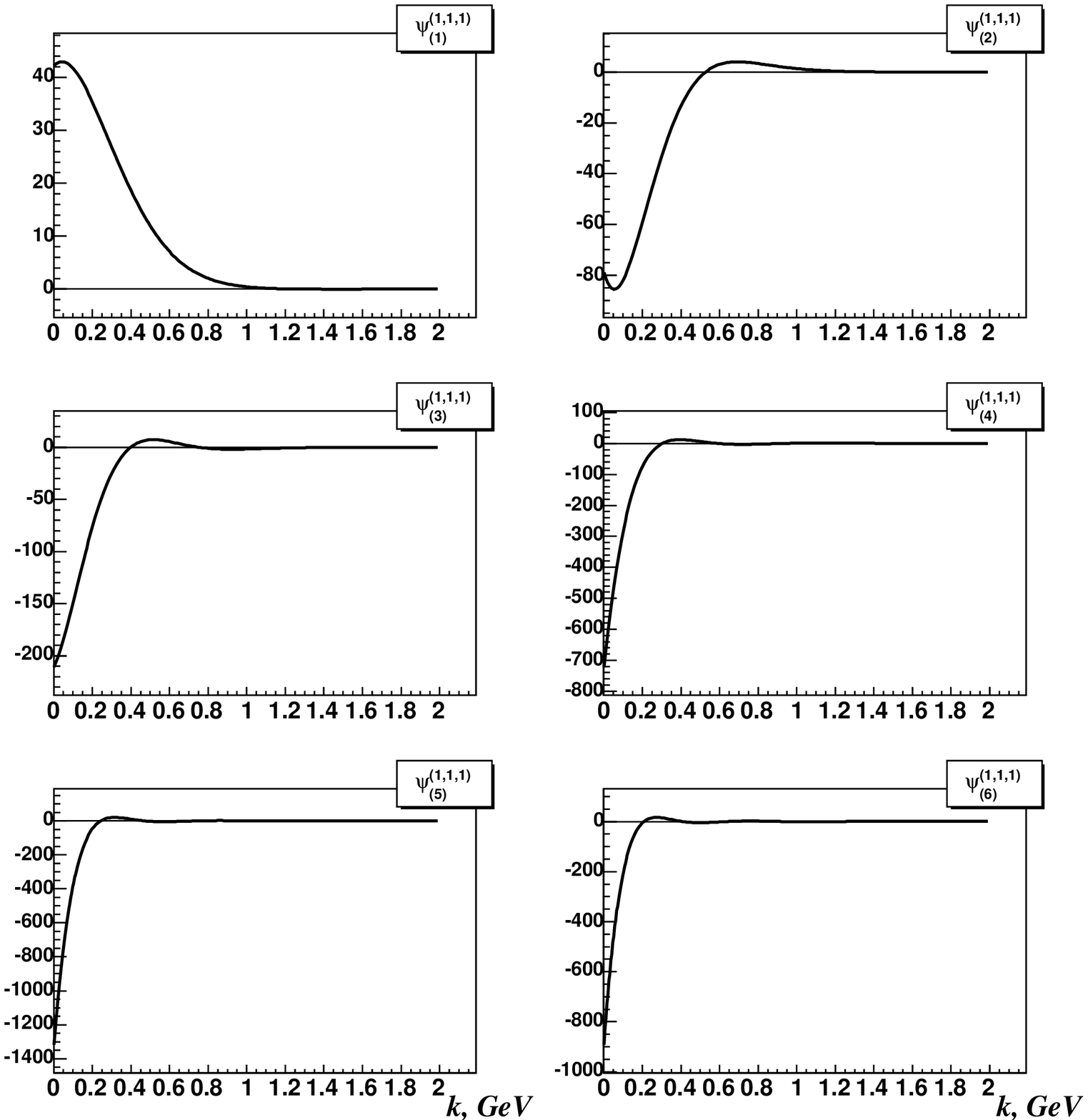,width=15cm}}
\caption{Wave functions of the  L=1 group
         ($a_1$-mesons with $n=1,2,3,4,5,6$).}
\end{figure}

\begin{figure}
\centerline{\epsfig{file=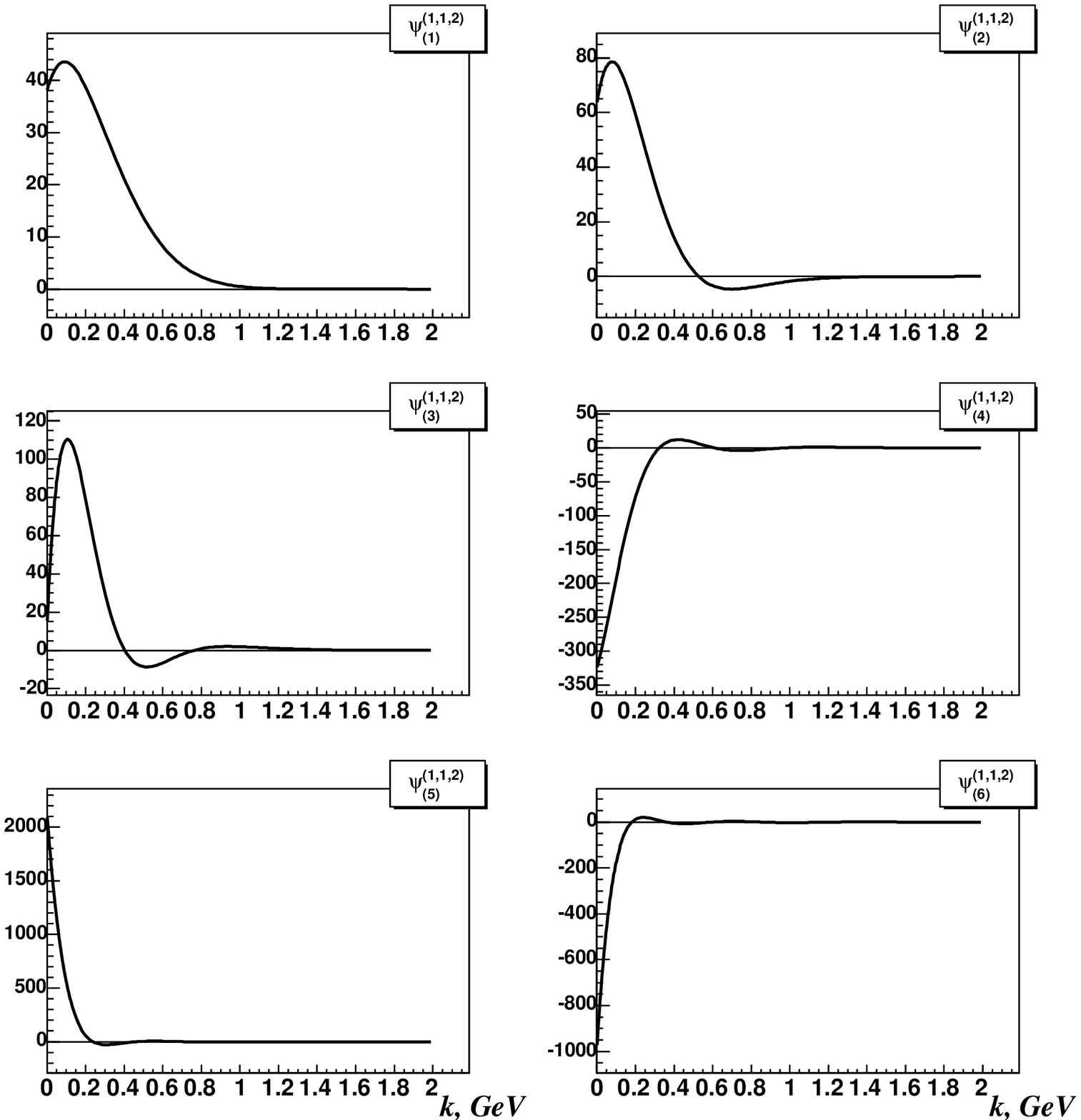,width=15cm}}
\caption{Wave functions of the  L=1 group
         ($a_2$-mesons with $n=1,2,3,4,5,6$).}
\end{figure}

\begin{figure}
\centerline{\epsfig{file=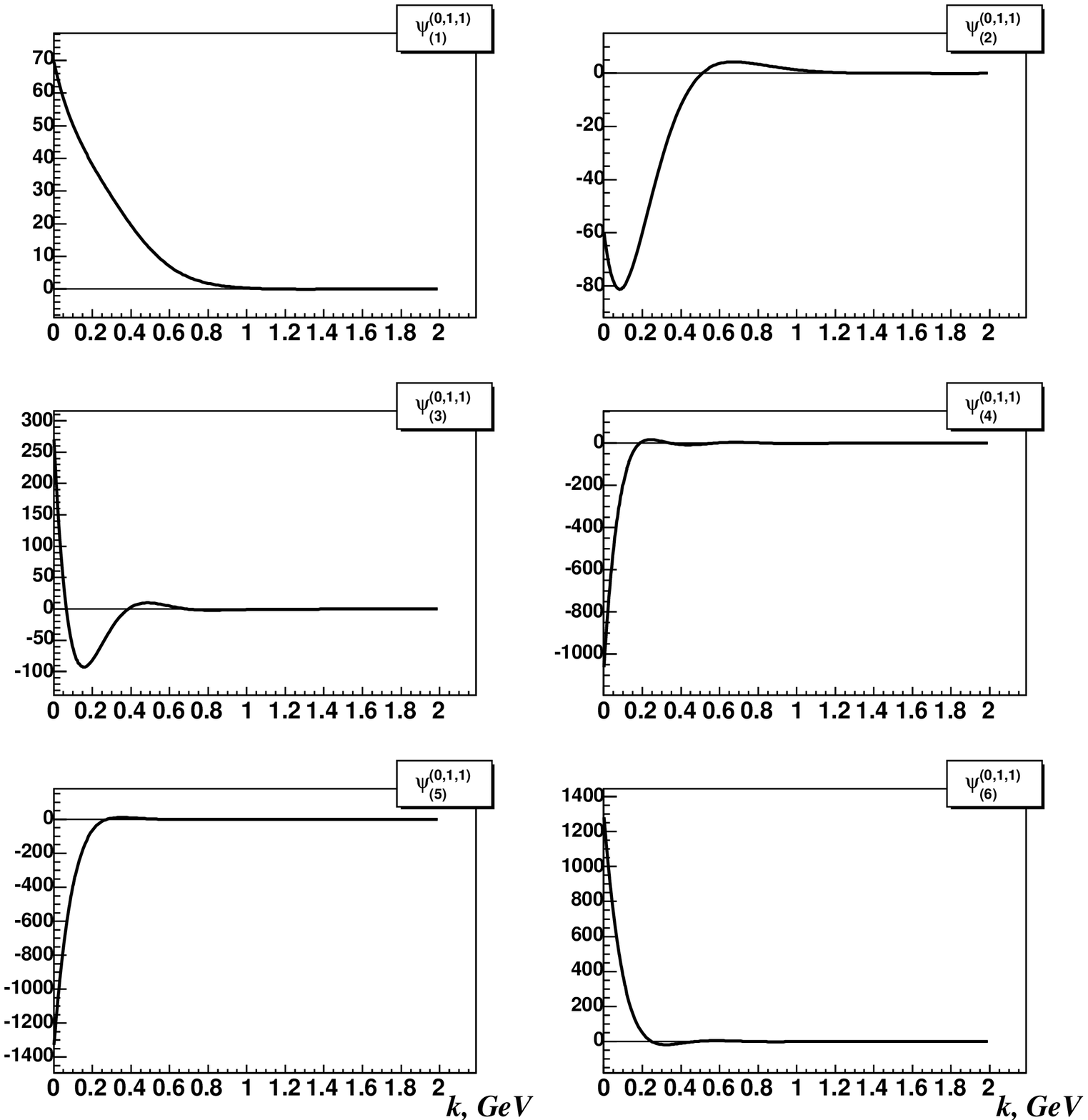,width=15cm}}
\caption{Wave functions of the  L=1 group
         ($b_1$-mesons with $n=1,2,3,4,5,6$).}
\end{figure}

\begin{figure}
\centerline{\epsfig{file=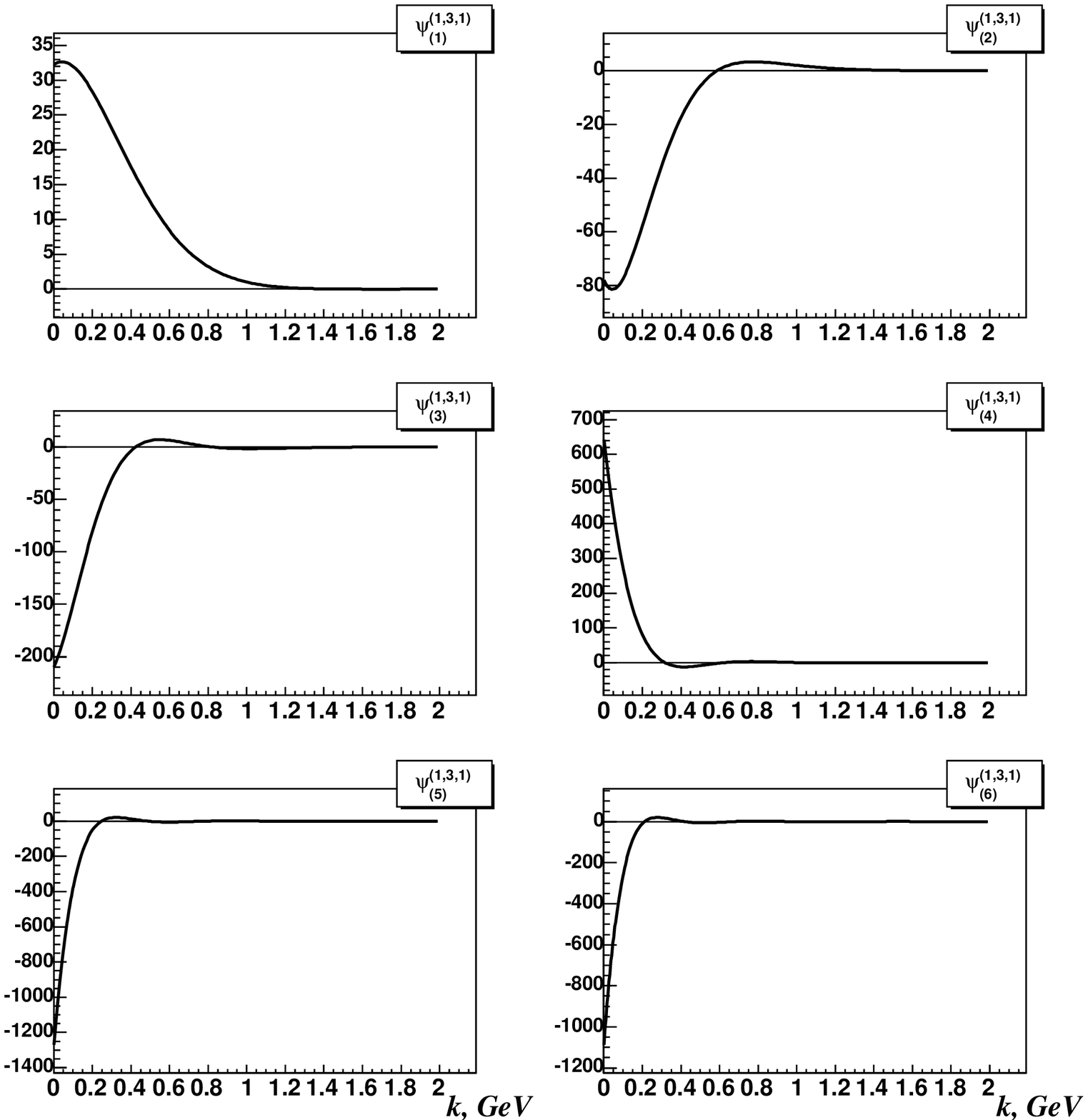,width=15cm}}
\caption{Wave functions of the  L=1 group
         ($f_2$-mesons with $n=1,2,3,4,5,6$).}
\end{figure}

\begin{figure}
\centerline{\epsfig{file=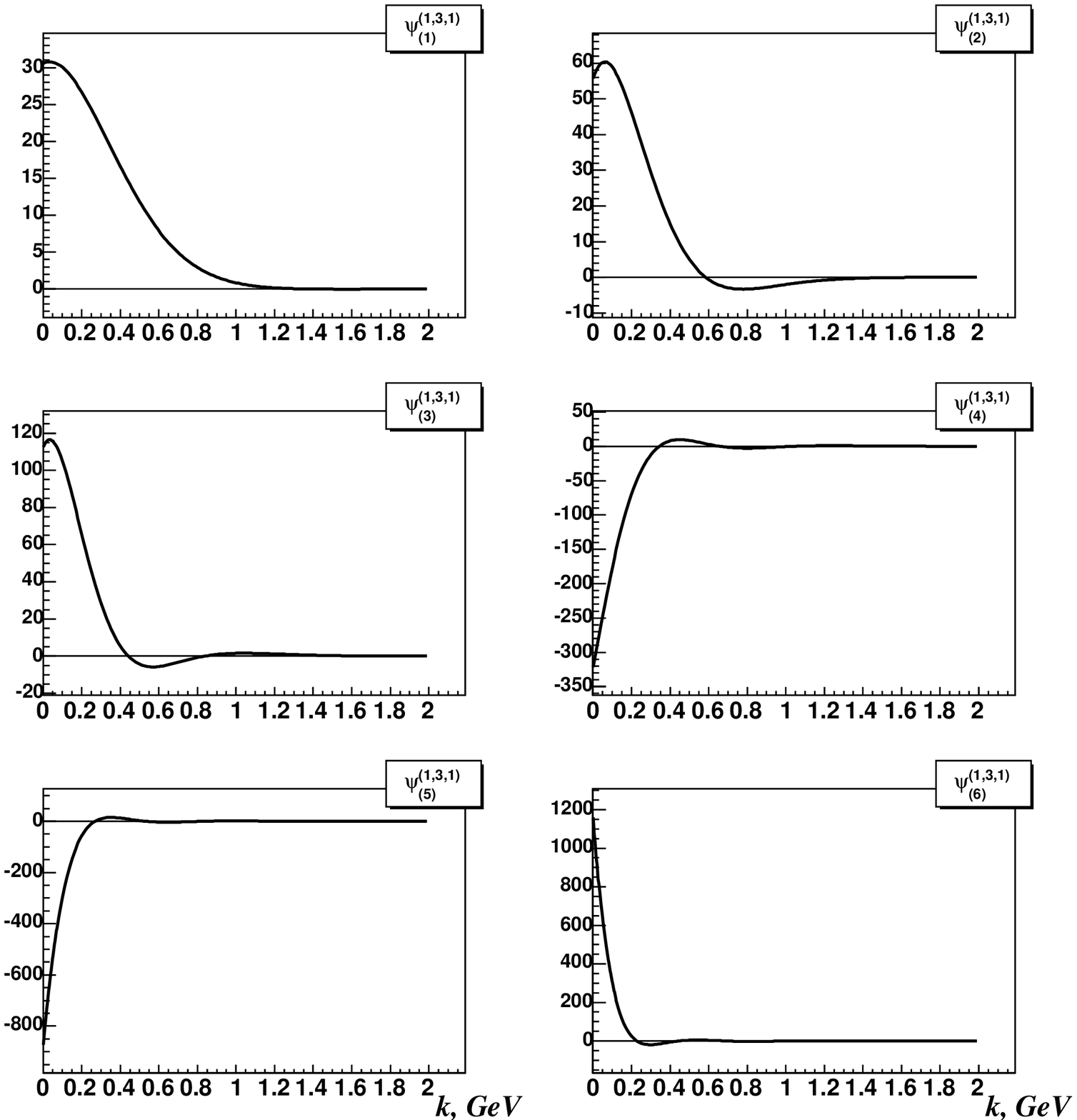,width=15cm}}
\caption{Wave functions of the  L=1 group
         ($f_2$-mesons with $n=1,2,3,4,5,6$).}
\end{figure}

\begin{figure}
\centerline{\epsfig{file=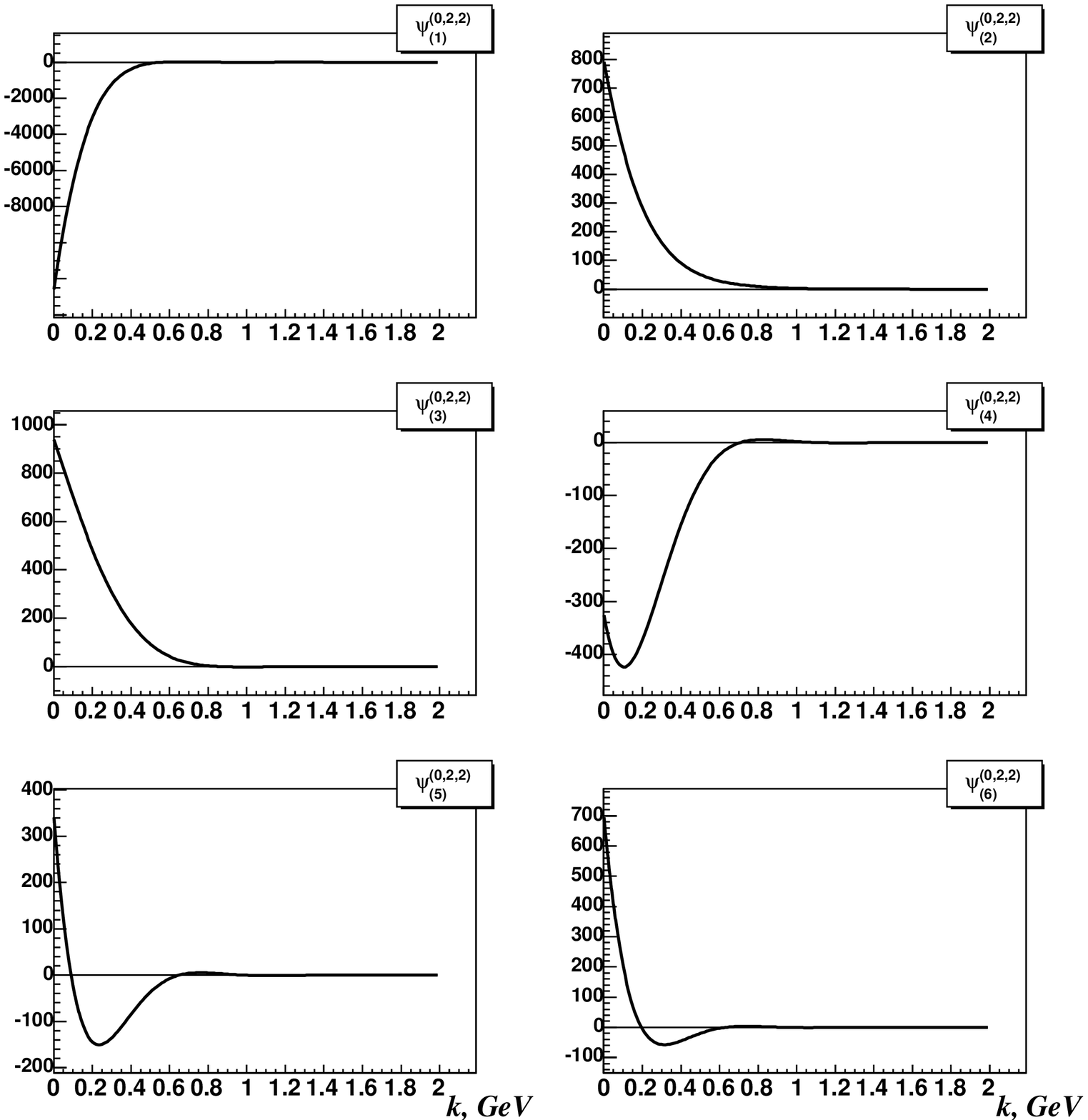,width=15cm}}
\caption{Wave functions of the  L=2 group
         ($\pi_2$-mesons with $n=1,2,3,4,5,6$).}
\end{figure}

\begin{figure}
\centerline{\epsfig{file=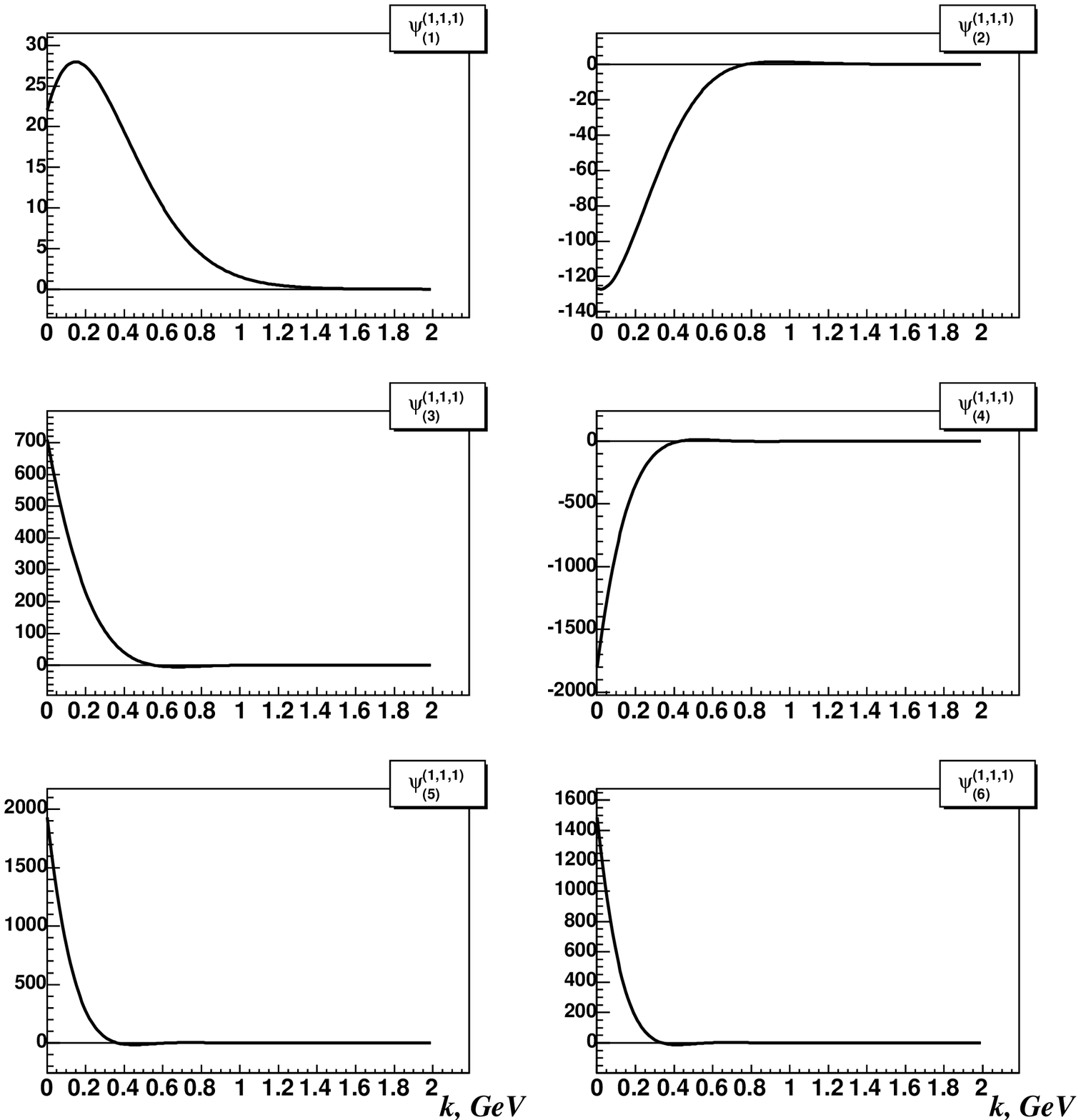,width=15cm}}
\caption{Wave functions of the  L=2 group
         ($\rho_1$- and $\omega_1$-mesons with $n=1,2,3,4,5,6$).}
\end{figure}

\begin{figure}
\centerline{\epsfig{file=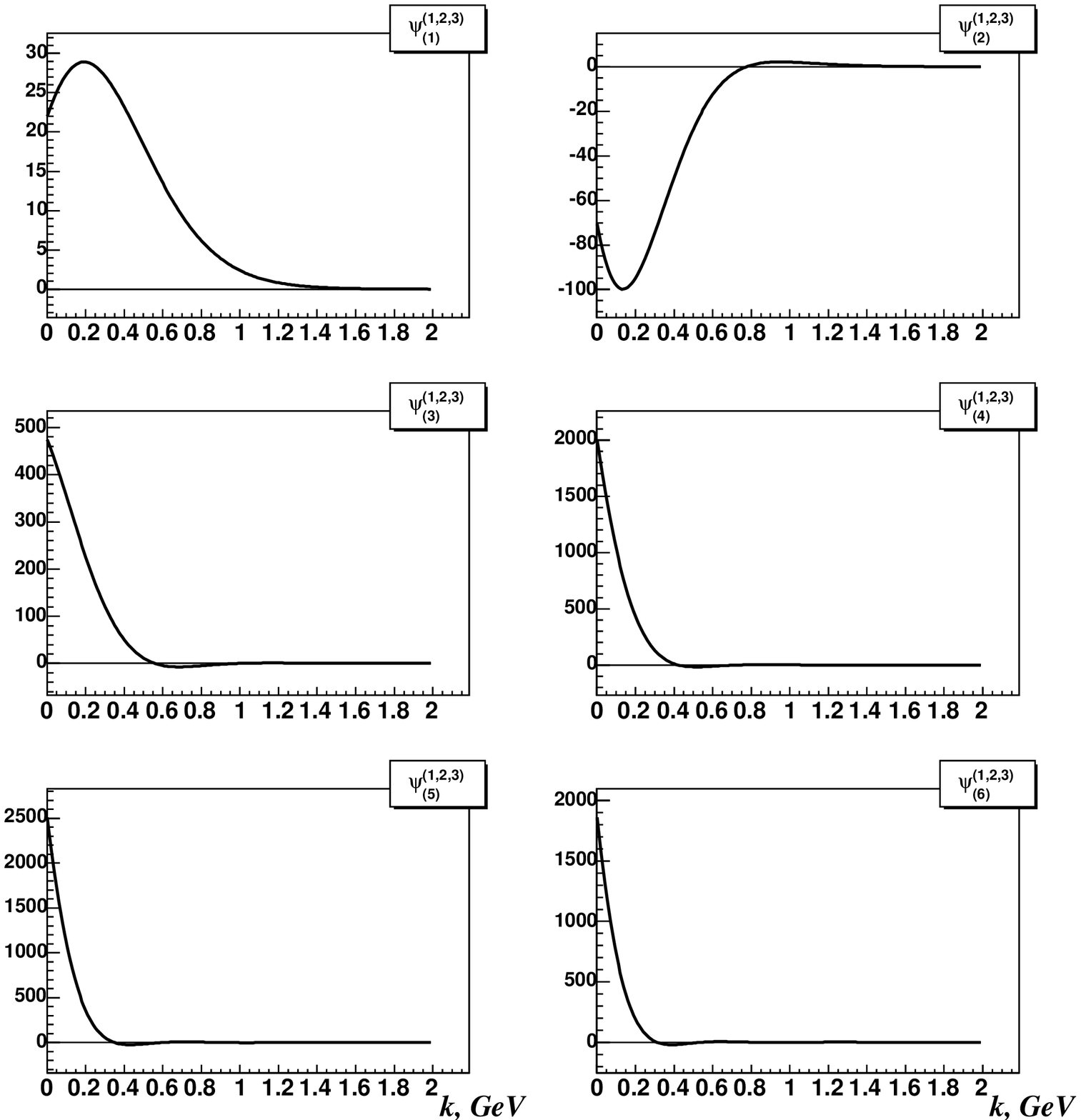,width=15cm}}
\caption{Wave functions of the  L=2 group
         ($\rho_3$- and $\omega_3$-mesons with $n=1,2,3,4,5,6$).}
\end{figure}

\begin{figure}
\centerline{\epsfig{file=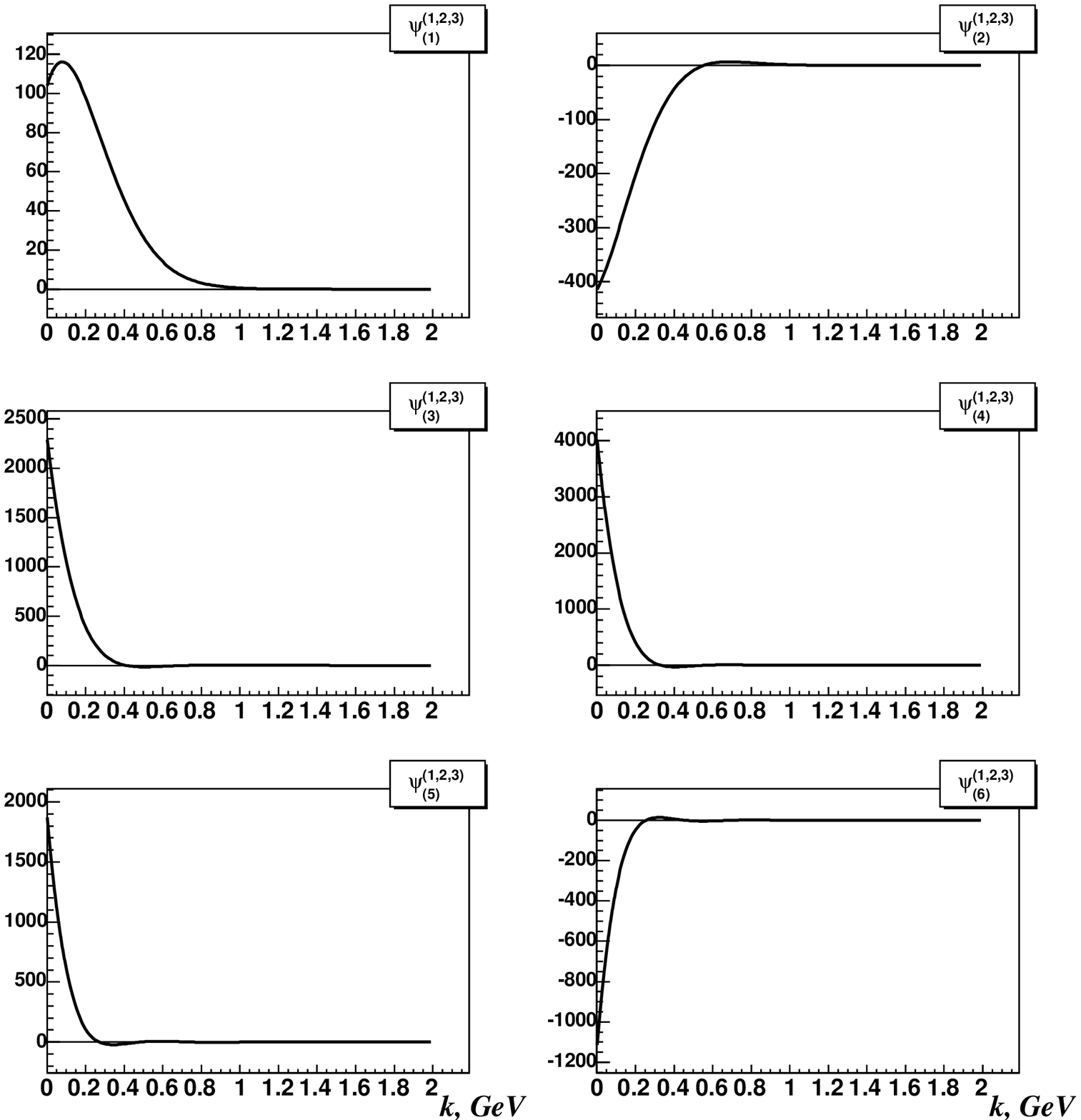,width=15cm}}
\caption{Wave functions of the  L=2 group
         ($\phi_3$-mesons with $n=1,2,3,4,5,6$).}
\end{figure}

\begin{figure}
\centerline{\epsfig{file=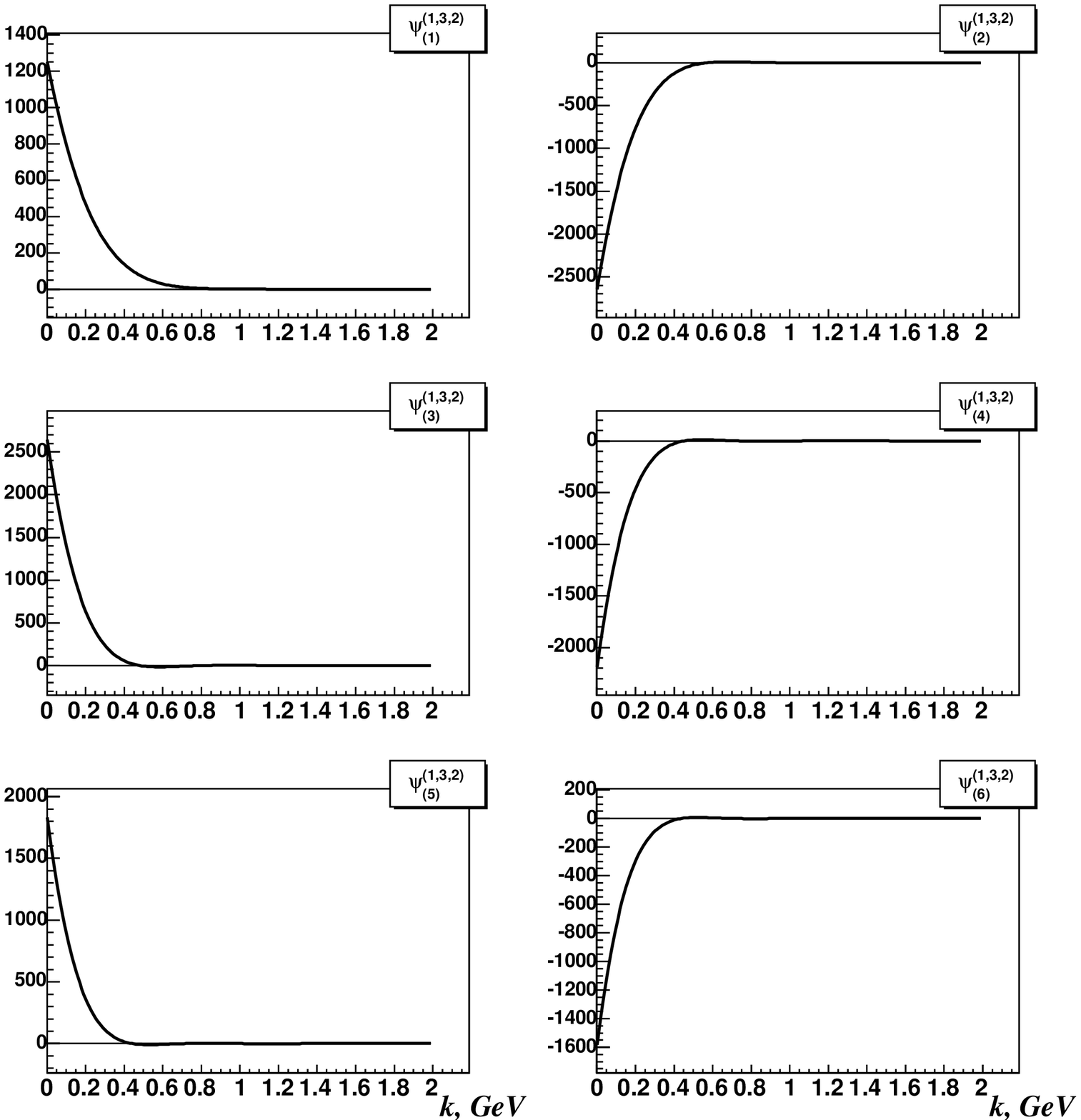,width=15cm}}
\caption{Wave functions of the  L=3 group
         ($a_2$-mesons with $n=1,2,3,4,5,6$).}
\end{figure}

\begin{figure}
\centerline{\epsfig{file=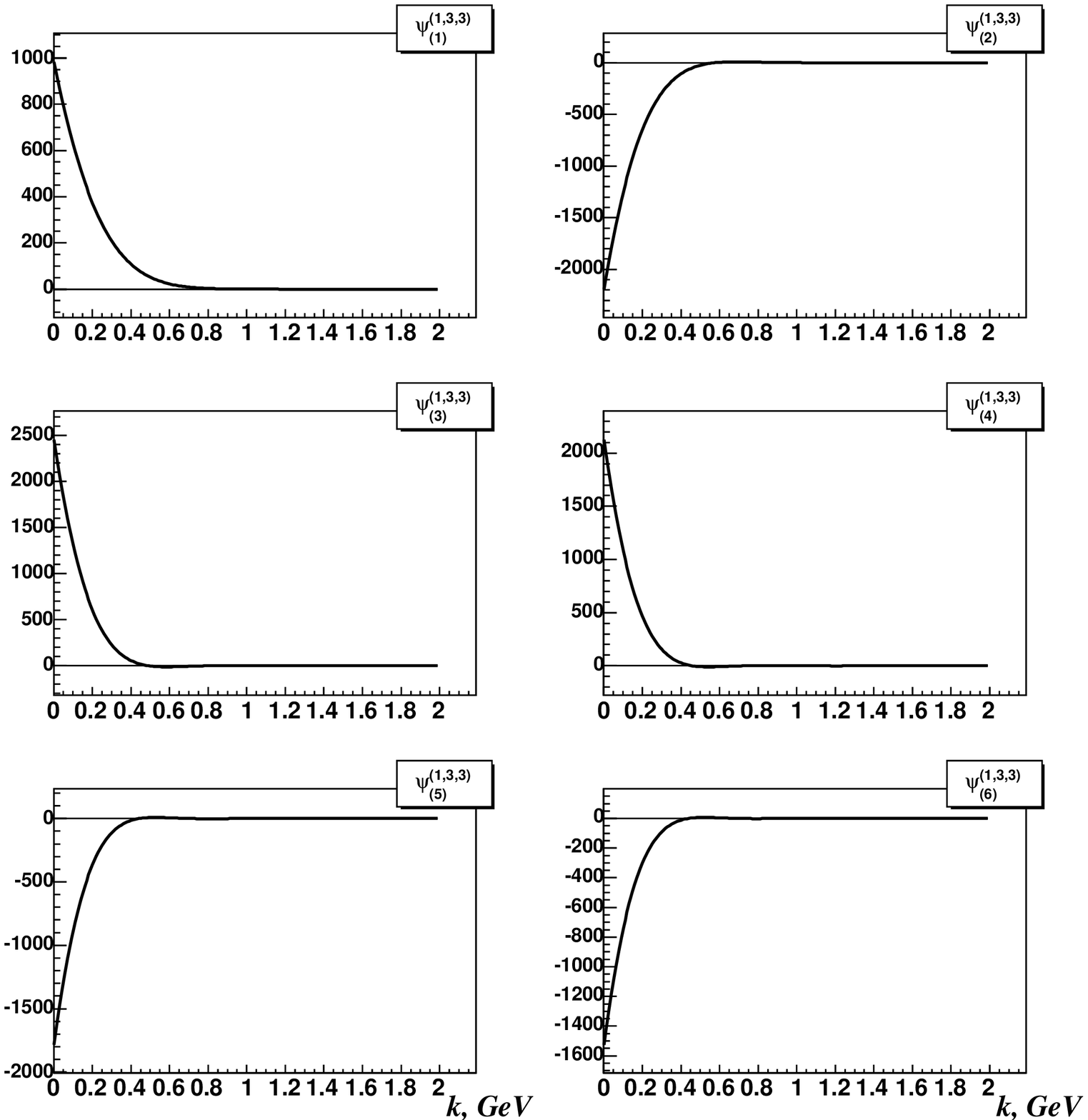,width=15cm}}
\caption{Wave functions of the  L=3 group
         ($a_3$-mesons with $n=1,2,3,4,5,6$).}
\end{figure}

\begin{figure}
\centerline{\epsfig{file=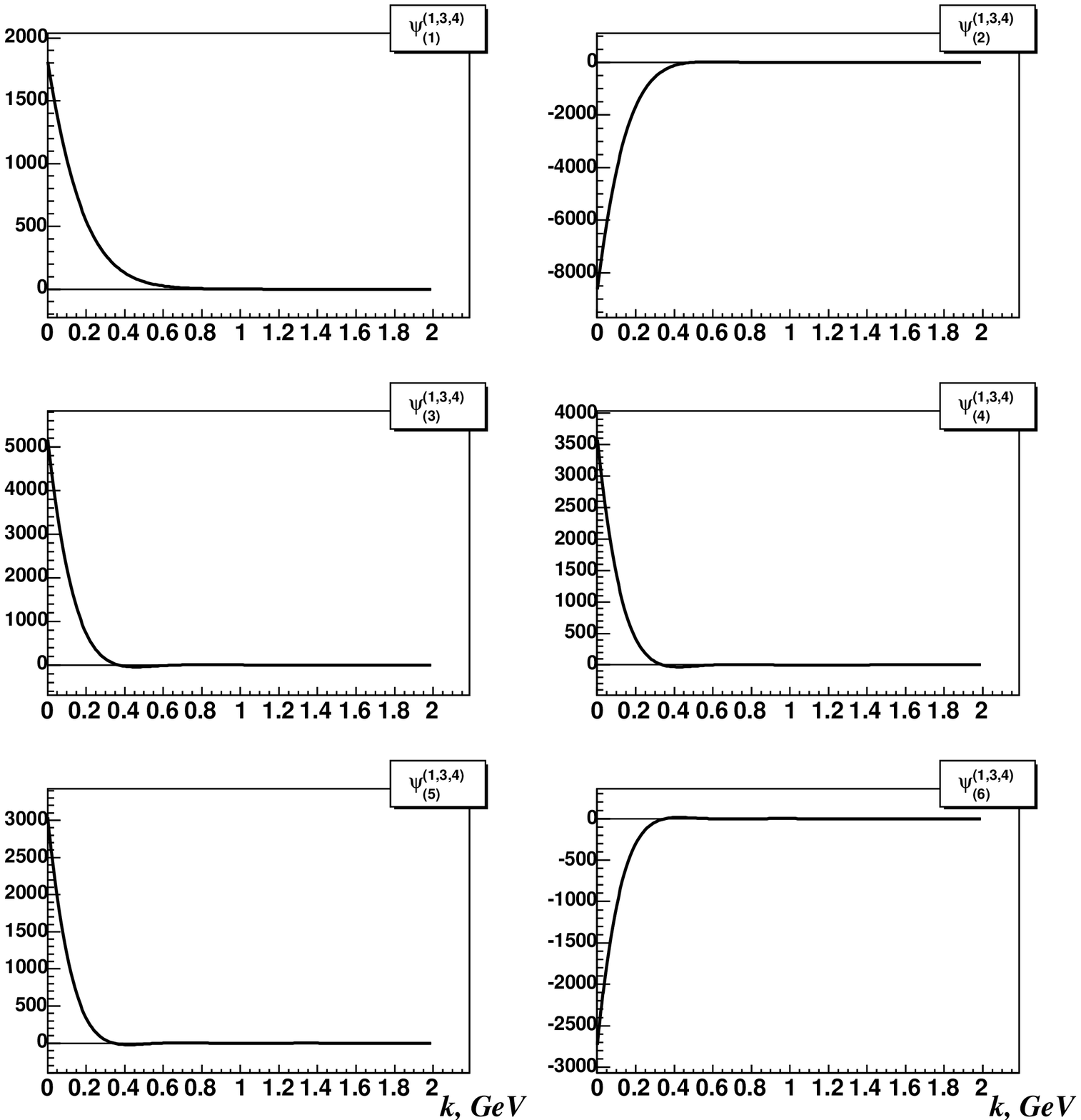,width=15cm}}
\caption{Wave functions of the  L=3 group
         ($a_4$-mesons with $n=1,2,3,4,5,6$).}
\end{figure}

\begin{figure}
\centerline{\epsfig{file=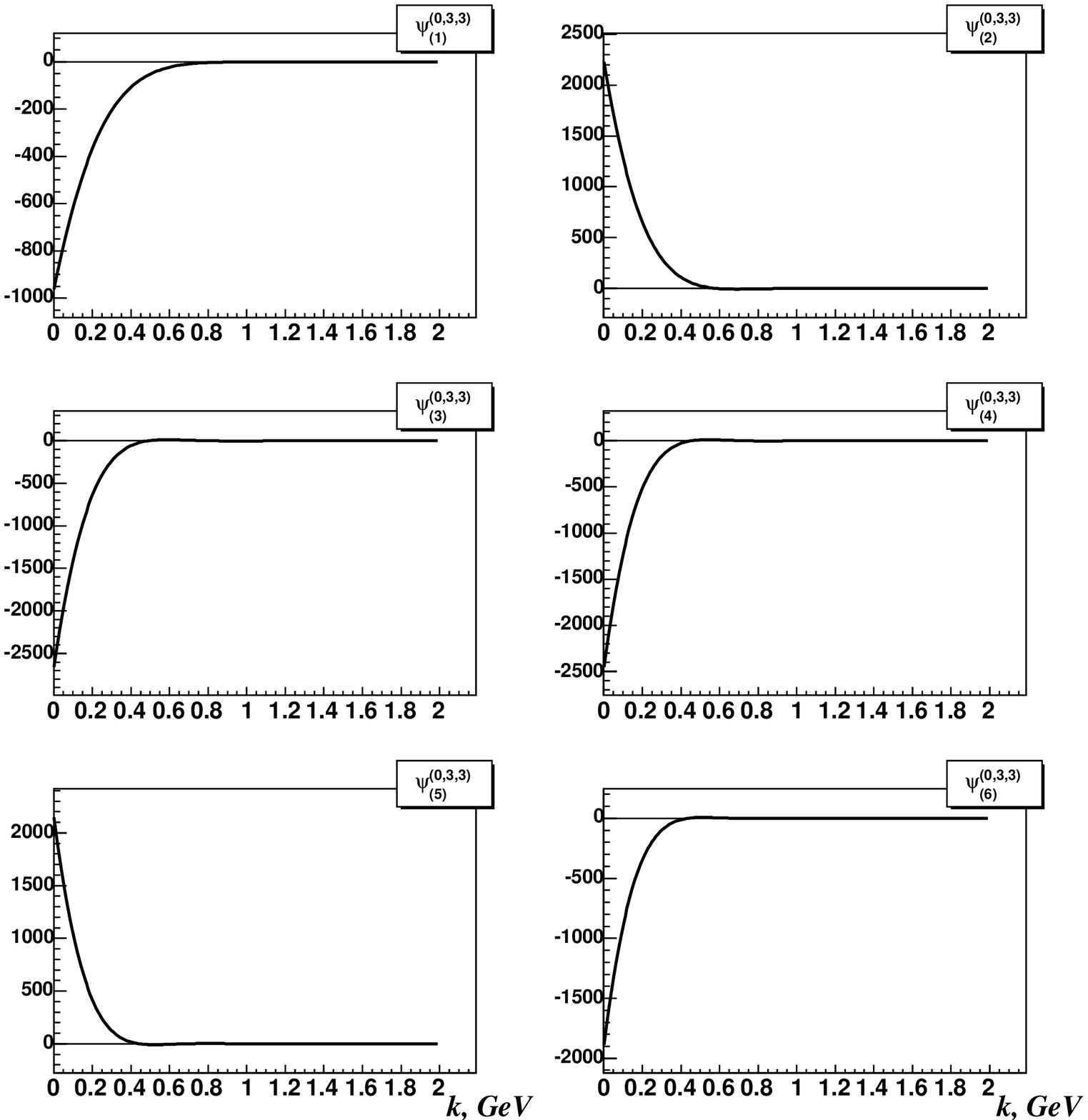,width=15cm}}
\caption{Wave functions of the  L=3 group
         ($b_3$-mesons with $n=1,2,3,4,5,6$).}
\end{figure}

\begin{figure}
\centerline{\epsfig{file=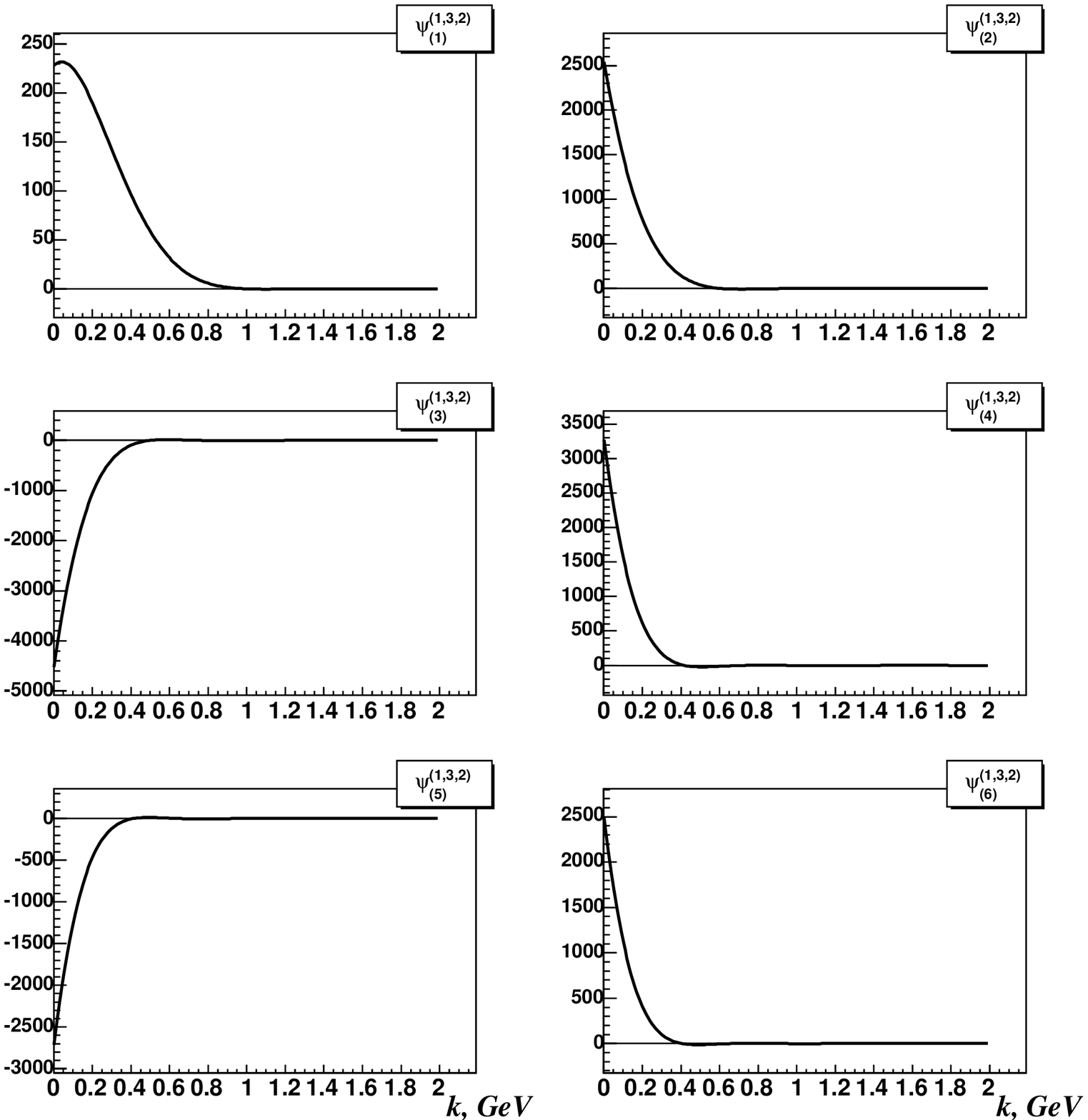,width=15cm}}
\caption{Wave functions of the  L=3 group
         ($f_2$-mesons with $n=1,2,3,4,5,6$).}
\end{figure}

\clearpage

\begin{figure}
\centerline{\epsfig{file=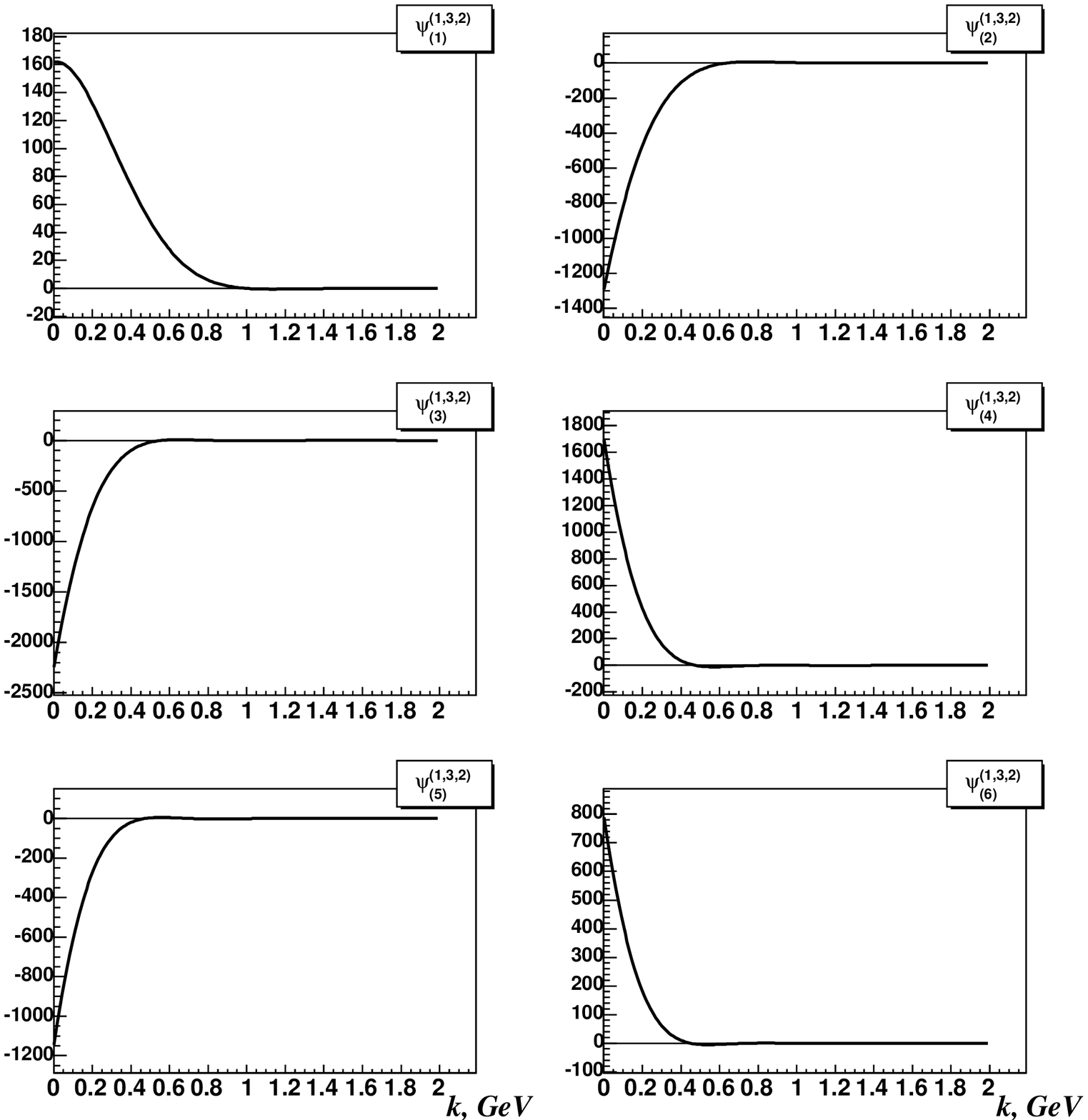,width=15cm}}
\caption{Wave functions of the  L=3 group
         ($f_2$-mesons with $n=1,2,3,4,5,6$).}
\end{figure}

\begin{figure}
\centerline{\epsfig{file=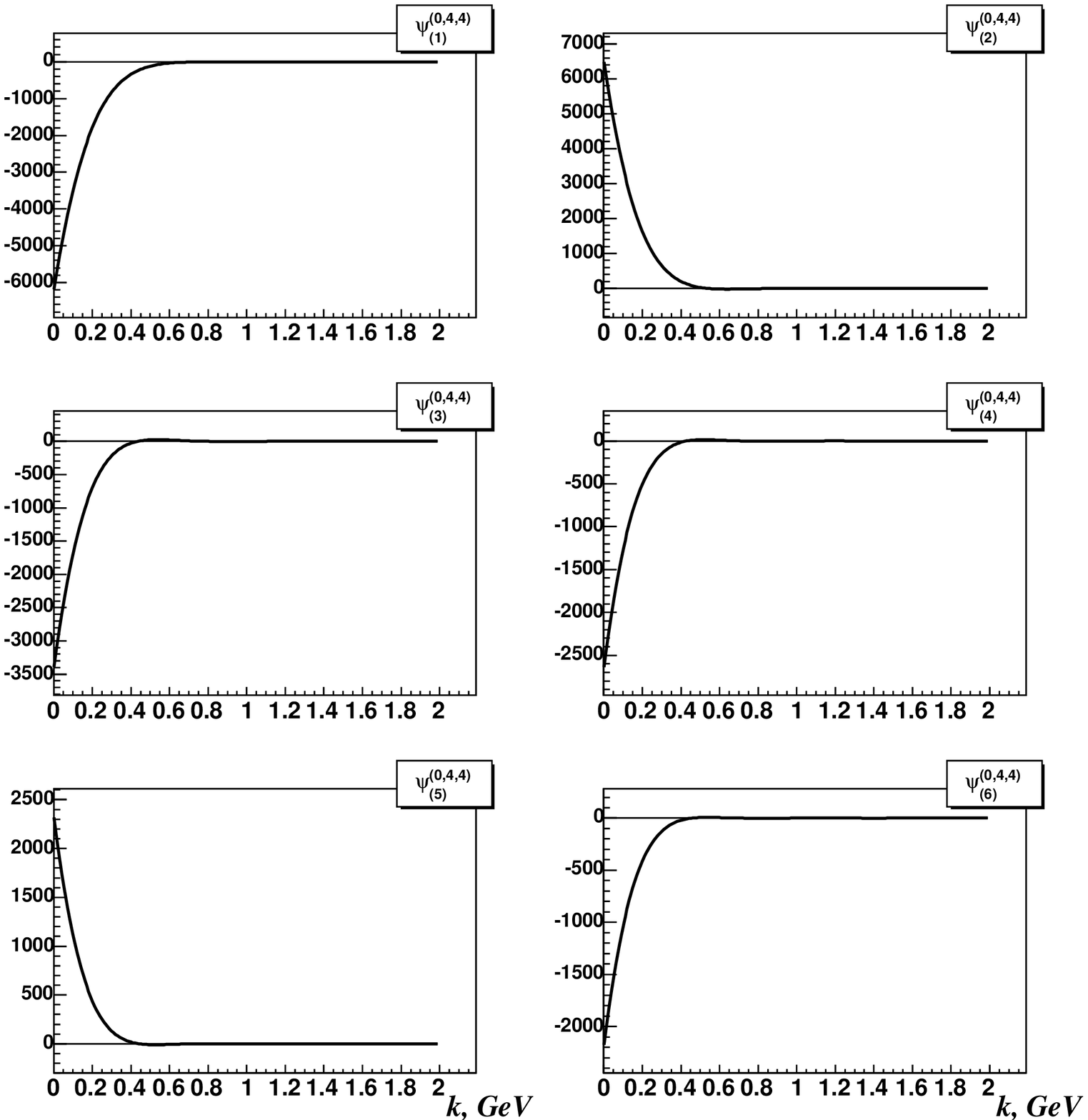,width=15cm}}
\caption{Wave functions of the  L=4 group
         ($\pi_4$-mesons with $n=1,2,3,4,5,6$).}
\end{figure}

\begin{figure}
\centerline{\epsfig{file=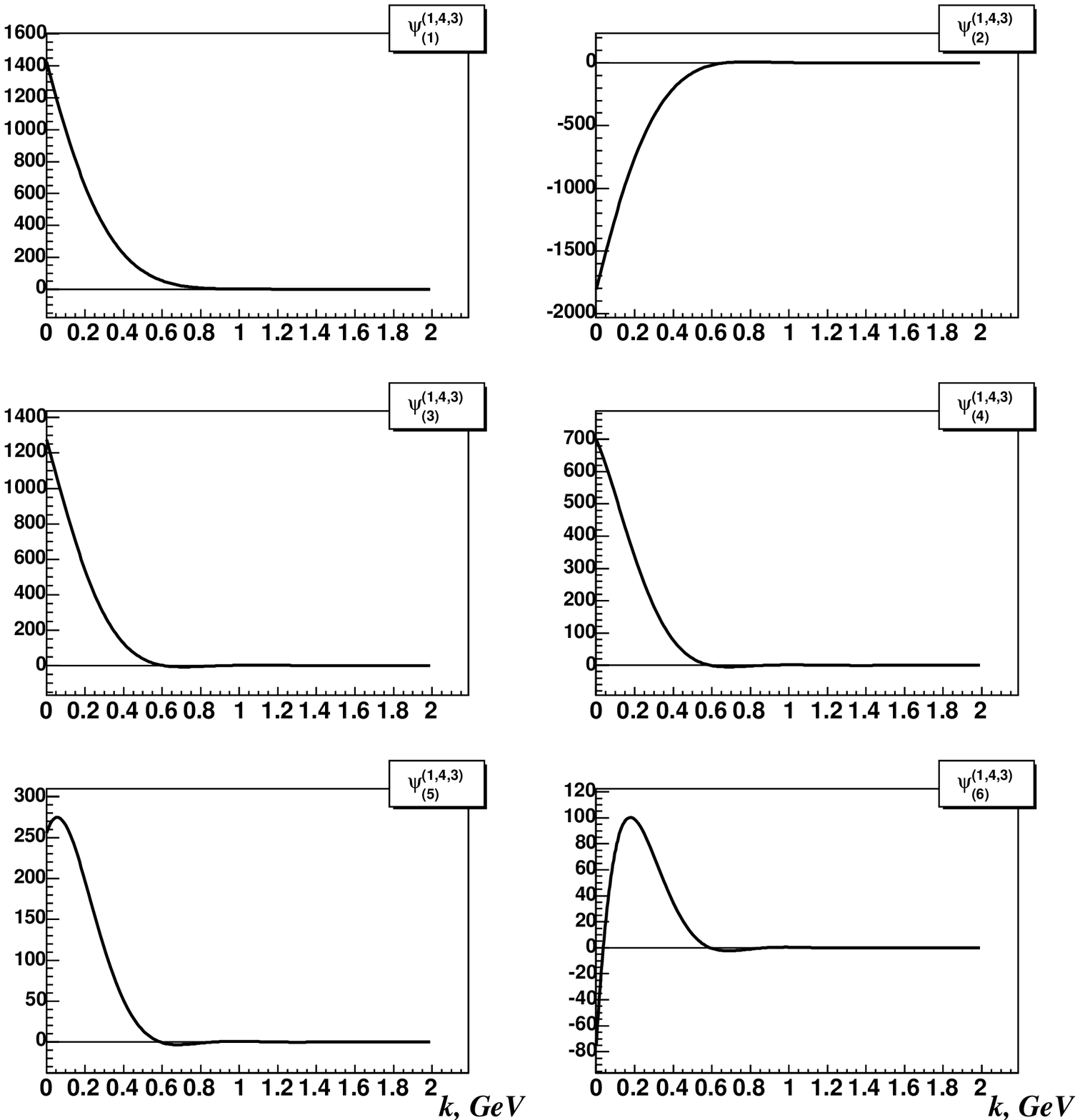,width=15cm}}
\caption{Wave functions of the  L=4 group
         ($\rho_3$-mesons with $n=1,2,3,4,5,6$).}
\end{figure}

\end{document}